\documentstyle{mymn}

\newif\ifAMStwofonts
\AMStwofontstrue

\ifoldfss
  \ifCUPmtlplainloaded \else
    \NewTextAlphabet{textbfit} {cmbxti10} {}
    \NewTextAlphabet{textbfss} {cmssbx10} {}
    \NewMathAlphabet{mathbfit} {cmbxti10} {} 
    \NewMathAlphabet{mathbfss} {cmssbx10} {} 
  \fi
  \ifAMStwofonts
    \ifCUPmtlplainloaded \else
      \NewSymbolFont{upmath} {eurm10}
      \NewSymbolFont{AMSa} {msam10}
      \NewMathSymbol{\upi}     {0}{upmath}{19}
      \NewMathSymbol{\umu}     {0}{upmath}{16}
      \NewMathSymbol{\upartial}{0}{upmath}{40}
      \NewMathSymbol{\leqslant}{3}{AMSa}{36}
      \NewMathSymbol{\geqslant}{3}{AMSa}{3E}

      \let\leq=\leqslant 
       
    \fi
  \fi
\fi 

\ifnfssone
  \newmathalphabet{\mathit}
  \addtoversion{normal}{\mathit}{cmr}{m}{it}
  \addtoversion{bold}{\mathit}{cmr}{bx}{it}
  \newmathalphabet{\mathbfit} 
  \addtoversion{normal}{\mathbfit}{cmr}{bx}{it}
  \addtoversion{bold}{\mathbfit}{cmr}{bx}{it}
  \newmathalphabet{\mathbfss} 
  \addtoversion{normal}{\mathbfss}{cmss}{bx}{n}
  \addtoversion{bold}{\mathbfss}{cmss}{bx}{n}
  \ifAMStwofonts
    \ifCUPmtlplainloaded \else
      \UseAMStwoboldmath
      \makeatletter
      \new@mathgroup\upmath@group
      \define@mathgroup\mv@normal\upmath@group{eur}{m}{n}
      \define@mathgroup\mv@bold\upmath@group{eur}{b}{n}
      \edef\UPM{\hexnumber\upmath@group}
      \new@mathgroup\amsa@group
      \define@mathgroup\mv@normal\amsa@group{msa}{m}{n}
      \define@mathgroup\mv@bold\amsa@group{msa}{m}{n}
      \edef\AMSa{\hexnumber\amsa@group}
      \makeatother
      \mathchardef\upi="0\UPM19
      \mathchardef\umu="0\UPM16
      \mathchardef\upartial="0\UPM40
      \mathchardef\leqslant="3\AMSa36
      \mathchardef\geqslant="3\AMSa3E

      \let\leq=\leqslant 

    \fi
  \fi
\fi 

\ifnfsstwo
  \DeclareMathAlphabet{\mathbfit}{OT1}{cmr}{bx}{it}
  \SetMathAlphabet\mathbfit{bold}{OT1}{cmr}{bx}{it}
  \DeclareMathAlphabet{\mathbfss}{OT1}{cmss}{bx}{n}
  \SetMathAlphabet\mathbfss{bold}{OT1}{cmss}{bx}{n}
  \ifAMStwofonts
    \ifCUPmtlplainloaded \else
      \DeclareSymbolFont{UPM}{U}{eur}{m}{n}
      \SetSymbolFont{UPM}{bold}{U}{eur}{b}{n}
      \DeclareSymbolFont{AMSa}{U}{msa}{m}{n}
      \DeclareMathSymbol{\upi}{0}{UPM}{"19}
      \DeclareMathSymbol{\umu}{0}{UPM}{"16}
      \DeclareMathSymbol{\upartial}{0}{UPM}{"40}
      \DeclareMathSymbol{\leqslant}{3}{AMSa}{"36}
      \DeclareMathSymbol{\geqslant}{3}{AMSa}{"3E}
      \DeclareMathSymbol{\gtrsim}       {\mathrel}{AMSa}{"26}
      \DeclareMathSymbol{\lesssim}      {\mathrel}{AMSa}{"2E}

      \let\leq=\leqslant 

    \fi
  \fi
\fi 

\ifCUPmtlplainloaded \else
  \ifAMStwofonts \else 
    \def\upi{\pi}
    \def\umu{\mu}
    \def\upartial{\partial}
  \fi
\fi

\title{High-resolution simulations of clump-clump collisions using SPH with Particle Splitting}

\author[S. Kitsionas \& A. P. Whitworth]
       {S. ~Kitsionas,$^{1,2,3}$\thanks{e-mail: skitsionas@aip.de} 
        A. P. ~Whitworth$^{3}$\thanks{e-mail: ant.whitworth@astro.cf.ac.uk}\\
$^{1}$Astrophysikalisches Institut Potsdam, An der Sternwarte 16, D-14482, Potsdam, Germany, \\
$^{2}$Institute of Astronomy \& Astrophysics, National Observatory of Athens, I. Metaxa and V. Pavlou, GR-15236, P. Penteli, Greece\\
$^{3}$School of Physics \& Astronomy, Cardiff University, P.O. Box 913,
5 The Parade, CF24 3AA, Cardiff, U.K.}

\date{Accepted.
      Received;
      in original form}

\pagerange{\pageref{firstpage}--\pageref{lastpage}}
\pubyear{2006}

\begin{document}

\maketitle

\label{firstpage}
	
\begin{abstract}
\noindent We investigate, by means of numerical simulations, the phenomenology 
of star formation triggered by low-velocity collisions between low-mass 
molecular clumps. The simulations are performed using an SPH code which 
satisfies the Jeans condition by invoking On-the-Fly Particle Splitting.

Clumps are modelled as stable truncated (non-singular) isothermal, i.e. 
Bonnor-Ebert, spheres. 
Collisions are characterised by $M_0$ (clump mass), $b$ (offset parameter, 
i.e. ratio of impact parameter to clump radius), and ${\cal M}$ (Mach Number, 
i.e. ratio of collision velocity to effective post-shock sound speed). 
The gas subscribes to a barotropic equation of state, which is intended to 
capture (i) the scaling of pre-collision internal velocity dispersion with 
clump mass, (ii) post-shock radiative cooling, and (iii) adiabatic heating 
in optically thick protostellar fragments.

The efficiency of star formation is found to vary between 10\% and 30\% in the 
different collisions studied and it appears to increase with decreasing $M_0$, 
and/or decreasing $b$, and/or increasing ${\cal M}$. For $b<0.5$ collisions 
produce shock compressed layers which fragment into filaments. Protostellar 
objects then condense out of the filaments and accrete from them. The 
resulting accretion rates are high, $1\;{\rm to}\;5 \times 10^{-5} 
M_\odot\,{\rm yr}^{-1}$, for the first $1\;{\rm to}\;3 \times 
10^4\,{\rm yrs}$. The densities in the filaments, $n_{{\rm H}_2} \gtrsim 5 
\times 10^5\,{\rm cm}^{-3}$, are sufficient that they could be mapped in 
NH$_3$ or CS line radiation, in nearby star formation regions.

\end{abstract}


\section{Introduction}{\label{sec:intro}}

There is increasing observational evidence that cloud-cloud collisions account 
for a substantial fraction of the star formation in the Galaxy 
\cite{ScovilleApJ1986,GreavesAnA1991,HasegawaApJ1994,Miyawaki1999,SatoApJ2000}. 
Recent theoretical calculations suggest that such collisions provide a 
viable mechanism for triggering star formation \cite{TanApJ2000}. Cloud-cloud 
collisions, like all dynamical star formation mechanisms, tend to result in 
the formation of groups of stars.

Because the observational evidence for star formation triggered by cloud-cloud 
collisions comes principally from relatively violent star formation events 
that spawn massive OB stars, most numerical work to date has been concerned 
with collisions between high-mass clouds. For example, the simulations of 
Chapman {\it et al}. \shortcite{ChapmanNATURE1992} treated collisions between 
$750 M_{\odot}$ clouds. However, if substructure in molecular clouds is 
hierarchical with a low volume-filling factor \cite{ScaloBLACK1985}, 
one might expect a collision between two high-mass clouds to consist of many 
smaller-scale collisions between the lower-mass clumps of which the clouds are 
composed. Such a picture presupposes that the clouds and clumps are 
long-lived equilibrium entities. However, even in the currently most popular 
paradigm of clouds being transient objects formed and destroyed by turbulent 
motions \cite{PadoanApJ2002,MacLowRMP2004,BerginApJ2004,VazquezApJ2005a}, 
fragmentation in shocks produced by large-scale converging flows can be 
studied in terms of smaller-scale shocks triggered by collisions between 
equilibrium clumps provided that such structures are of sufficiently low-mass.

The advantage of simulating collisions between lower-mass clumps is that the 
resolution requirements are less severe. It has been shown, both in the 
context of Finite Difference simulations 
\cite{TrueloveApJ1997,TrueloveApJ1998}, and in the 
context of Smoothed Particle Hydrodynamics (SPH) simulations 
\cite{BateMNRAS1997,KitsionasMNRAS2002}, that gravitational fragmentation can 
only be modelled faithfully if the Jeans mass is resolved at all times. This 
requirement is normally referred to as the Jeans condition. Previous 
simulations of clump-clump collisions 
\cite{ChapmanNATURE1992,PongracicMNRAS1992,TurnerMNRAS1995,WhitworthMNRAS1995,BhattalMNRAS1998,MarinhoAnAS2000,MarinhoAnA2001}
followed the evolution through several orders of magnitude in density and 
linear scale, and produced abundant high-mass protostellar fragments by 
gravitational fragmentation. However, these simulations did not satisfy the 
Jeans condition. Therefore, it is possible that real fragmentation on small 
scales was suppressed \cite{WhitworthMNRAS1998a}, and/or that artificial 
fragmentation occurred. In this paper, we describe high-resolution simulations 
of clump-clump collisions, performed using SPH with On-the-Fly Particle 
Splitting \cite{KitsionasMNRAS2002}. Particle Splitting enables us to ensure 
that the Jeans condition is satisfied at all times. This is achieved with 
relatively modest computing resources, by 
introducing high resolution only in the regions where it is required.

Two suites of simulations are presented aiming at i) the identification, 
free from numerical resolution constraints, of the dominant mechanism driving 
star formation in clump collisions and ii) the derivation of, at least, order 
of magnitude estimates for the star formation efficiency in such collisions. 
In the first (and principal) suite of simulations, 
collisions between low-mass clumps having mass $M_0 = 10 M_\odot$ are 
simulated with various combinations of impact parameter and collision 
velocity. In the second suite, a few collisions between intermediate-mass 
clumps having $M_0 = 75 M_\odot$ 
are simulated. The purpose of the second suite is twofold: (i) to explore the 
dependence on clump mass, and (ii) to repeat some of the critical simulations 
reported by Bhattal {\it et al}. \shortcite{BhattalMNRAS1998} and establish 
which of the features they inferred might be attributable to inadequate 
resolution.

In Section \ref{sec:model}, we describe the physical model we use. In 
Section \ref{sec:sph} we give a brief summary of our SPH code and explain how 
Particle Splitting is invoked to ensure the Jeans condition is always 
satisfied. A representative selection from the main suite of simulations 
(those using $10 M_{\odot}$ clumps) is presented in Section 
\ref{sec:low-mass}. Results from the second suite ($75 M_\odot$ clumps) are 
presented in Section \ref{sec:amar} and compared with the results of Bhattal 
{\it et al} \shortcite{BhattalMNRAS1998}. In Section \ref{sec:conclusi} 
we discuss the results and summarise our main conclusions.  

\section{Physical Model}{\label{sec:model}}

\subsection{Clump and collision parameters}

In all cases, the collision involves two clumps of equal mass. In molecular 
clouds with hierarchical substructure, collisions between clumps at the same 
level of the hierarchy (i.e. with comparable mass) are the most probable 
\cite{ScaloBLACK1985} -- although it would certainly be interesting to 
investigate collisions involving clumps of unequal mass.

The collision is set up in the centre-of-mass frame, and so the clumps have 
antiparallel bulk velocities ${\bf v}_{\rm clump}$ and 
$-\,{\bf v}_{\rm clump}$. 
We define the Mach Number (${\cal M}$) of the collision as the ratio of the 
relative speed of the collision ($2\,|{\bf v}_{\rm clump}|$) to the 
effective post-shock sound speed ($c_{\rm s}$, see Eqn. \ref{equa:statecool} 
below). We define the offset parameter $b$ as the ratio of the impact 
parameter of the collision to the clump radius.

We take the clumps to be stable equilibrium isothermal spheres, 
i.e. they are contained by an external pressure and they are not singular 
\cite{EbertZA1955,BonnorMNRAS1956}. 
In order to create such clumps we apply the procedure detailed in 
Appendix A4 of Turner {\it et al}. \shortcite{TurnerMNRAS1995}. As 
demonstrated in Appendix B2 of Turner {\it et al}. \shortcite{TurnerMNRAS1995} 
this reproduces faithfully the density profile of a stable 
isothermal sphere, with very little particle noise. 

In order to save computational time, the colliding clumps are touching when 
the simulations start. During the preceding approach, mutual tidal distortion 
should be small because the clumps move supersonically 
($\mathcal{M} \gtrsim$ 5). We
define the $x$-axis to be parallel to the pre-collision velocities, and the 
$y$-axis to be parallel to the impact parameter. Hence the global angular 
momentum of the colliding clumps is parallel to the $z$-axis.

In the context of a turbulent interstellar medium, the above model can be 
interpreted as the interaction of flows generated at the scale of the 
equilibrium clumps employed here either by sources driving turbulence at these 
scales or by the energy cascade of turbulence driven at larger 
scales \cite{PadoanApJ2002,MacLowRMP2004,BerginApJ2004,VazquezApJ2005b}. 
The modelling of turbulent motions on smaller scales is 
not included self-consistently here, but rather it is 
represented by
an isotropic pressure force in the effective 
sound speed (Eqn. \ref{equa:statecool}) of the clumps, in accordance with the 
scaling relations of Larson \shortcite{LarsonMNRAS1981} between the internal 
velocity dispersion and the linear scale or the mass of a clump (e.g. see Eqn. 
\ref{equa:larson}). The use of an effective 
turbulent pressure to model turbulence acting on small scales suppresses the 
stochastic nature of any real density and velocity fluctuations expected from 
the presence of (supersonic or even transsonic) turbulent motions inside the 
clumps. Such an approach takes into account only the collective effect such 
fluctuations may have on the stability of the clump as a whole, neglecting the 
fact that small scale fluctuations can in turn induce collapse on these scales 
\cite{MacLowRMP2004}, as we mainly want to identify the role of the collision 
itself on the triggering of star formation.

Moreover, the current model does not take into account the effect that 
magnetic fields may have on the dynamics and the final outcome of the 
collisions. Although it is known that almost all studied clumps and cores 
are seen to be close to magnetic criticality \cite{CrutcherApJ2004}, we 
effectively study here collisions between clumps threaded by weak magnetic 
fields, which will not become dynamically important during the course of 
the collisions. Given the fact that i) collisions between low-mass clumps have 
not been studied before and ii) the numerical methods we employ provide an 
adequate framework for studying self-gravitating hydrodynamics at the highest 
numerical resolution necessary, we believe that our assumption of magnetic 
subcriticality for the clumps is a reasonable first approach to this problem.

\begin{figure}

\setlength{\unitlength}{1mm}
\begin{picture}(80,80)
\includegraphics{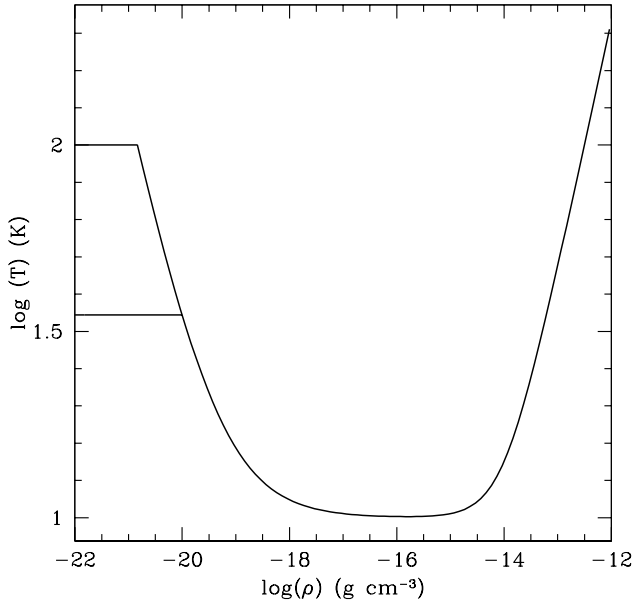}
\end{picture}

\caption{The equation of state given in Eqn. \ref{equa:statecool} presented as 
a density-temperature relation with $c_{0}$ given from Eqn. \ref{equa:larson} 
for the initial clump masses adopted here (see next session), 
$c_{\rm s} \simeq 0.2\,{\rm km}\,{\rm s}^{-1}$, $\rho_0 = \rho_c$ 
($\rho_{\rm c}$ is the initial central density of a clump; see next section), 
$\rho_1 \simeq 10^{-14}\,{\rm g}\,{\rm cm}^{-3}$. As soon as the collision 
starts the gas cools with $T \propto \rho^{-2/3}$ down to 
$T \sim 10\,{\rm K}$, a temperature that is reached at 
$\sim 10^{-18}\,{\rm g}\,{\rm cm}^{-3}$. Then it remains isothermal at 
$T = 10\,{\rm K}$, as it is thermally coupled to the dust, until it gets 
optically thick to its own cooling radiation and heats up adiabatically with 
$T \propto \rho^{2/3}$, for densities above 
$10^{-14}\,{\rm g}\,{\rm cm}^{-3}$.}
\label{fig:eos}
\end{figure}

\subsection{Equation of state}{\label{sec:eos}}

We use a barotropic equation of state, 

\begin{equation}
\label{equa:statecool}
\frac{P}{\rho} \equiv c^2 = \left\{ \begin{array}{lr}
c_0^2\,, & \rho \leq \rho_0\,; \\
\, & \, \\
\left[ (c_0^2 - c_{\rm s}^2)(\rho/\rho_0)^{-2/3} + c_{\rm s}^2 \right] & \\
\hspace{0.6cm} \times \left[ 1 + (\rho/\rho_1)^{4/3} \right]^{1/2}\,, & \rho > \rho_0\,, \\
\end{array} \right.
\end{equation}
\noindent that is illustrated in Fig. \ref{fig:eos}. 
The left hand side 
represents the square of the effective isothermal sound speed 
of the gas, i.e. when non-thermal pressure due to turbulence is included.

At low densities, $\rho \leqslant \rho_0$ (i.e. before the collision) the 
clump gas has an effective sound speed
\begin{equation}
\label{equa:larson}
c_{0} \;\sim\; 0.2\,{\rm km}\,{\rm s}^{-1} \left( \frac{M_{0}}{M_{\odot}} 
\right)^{1/4} 
\end{equation}
in accordance with Larson's scaling relation between the mass and internal 
velocity dispersion of a clump \cite{LarsonMNRAS1981}.

Once the density exceeds $\rho_0 = \rho_{\rm c}$ ($\rho_{\rm c}$ is the 
initial central density of a clump; see the next section for the specific 
values adopted here), i.e. essentially as soon as the gas is compressed by the 
collision shock, 
the gas cools down (with $T \propto \rho^{-2/3}$, see the first term on 
the right hand side of Eqn. \ref{equa:statecool} for $\rho > \rho_0$) 
to $T \sim 10\,{\rm K}$ 
\cite{WhitworthMNRAS1997,WhitworthMNRAS1998b}. The post-shock sound speed 
is therefore $c_{\rm s} \simeq 0.2\,{\rm km}\,{\rm s}^{-1}$ (corresponding 
to a cosmic mixture of H$_2$ and He at 10 K). The gas reaches for the 
first time this temperature, and thus this sound speed, 
at $\sim 10^{-18}\,{\rm g}\,{\rm cm}^{-3}$ and then it evolves isothermally 
for a few orders of magnitude in density (see Fig. \ref{fig:eos}). 

This part of the equation of state is similar to the piecewise 
polytropic equation of state advocated by Larson \shortcite{LarsonMNRAS2005}. 
It differs mainly in the polytropic exponent assumed for the regime in which 
the gas is thermally coupled to the dust, i.e. above 
$\sim 10^{-18}\,{\rm g}\,{\rm cm}^{-3}$ 
\cite{WhitworthMNRAS1997,WhitworthMNRAS1998b}. For this regime, 
Larson's equation of state assumes slow heating of the gas 
($\gamma \simeq 1.1$), whereas we have taken here the 
gas to evolve isothermally ($\gamma = 1$).

Finally, once the density in a collapsing protostellar object exceeds 
$\rho_1 \simeq 10^{-14}\,{\rm g}\,{\rm cm}^{-3}$, the gas is expected to 
become sufficiently optically thick to its own cooling radiation that it heats 
up adiabatically with $T \propto \rho^{2/3}$ \cite{TohlineCANUTO1982}, 
given by the second term on the right hand side of Eqn. 
\ref{equa:statecool} for $\rho > \rho_0$. We note that for a cosmic 
mixture of H$_2$ and He at temperatures below 500 K, the adiabatic exponent 
is $\gamma \simeq 5/3$ because the rotational degrees of freedom of H$_2$ are 
frozen out; in our simulations the temperature does not rise above 200 K.

\begin{table*}
\begin{center}
\begin{tabular}{ccccccccccc}
$b$ & $\mathcal{M}$ & $v_{\rm coll}$ & $v_{\rm clump}$ & $M_{0}$ & $c_{0}$ & 
$T$ & $t_{\rm frag}$ & $M_{\rm frag}$ & $t_{\rm end}$ & 
$t_{\rm evol}$ \\ \hline 
0.0 & 5  & 1.0 & 0.5 & 10 & 0.35 & 35  & 0.472 & 0.33 & 0.476 & 1.08 \\ 
0.2 & 5  & 1.0 & 0.5 & 10 & 0.35 & 35  & 0.479 & 0.70 & 0.496 & 1.34 \\ 
0.4 & 5  & 1.0 & 0.5 & 10 & 0.35 & 35  & 0.535 & 1.11 & 0.557 & 1.44 \\ 
0.6 & 5  & 1.0 & 0.5 & 10 & 0.35 & 35  & 0.678 & 0.68 & 0.701 & 1.46 \\ 
0.8 & 5  & 1.0 & 0.5 & 10 & 0.35 & 35  & --    & --   & --    & --   \\ \hline
0.0 & 10 & 2.0 & 1.0 & 10 & 0.35 & 35  & 0.360 & 0.35 & 0.370 & 1.20 \\ 
0.2 & 10 & 2.0 & 1.0 & 10 & 0.35 & 35  & 0.360 & 1.00 & 0.396 & 1.72 \\ 
0.4 & 10 & 2.0 & 1.0 & 10 & 0.35 & 35  & 0.485 & 0.48 & 0.507 & 1.44 \\ 
0.6 & 10 & 2.0 & 1.0 & 10 & 0.35 & 35  & --    & --   & --    & --   \\ 
0.8 & 10 & 2.0 & 1.0 & 10 & 0.35 & 35  & --    & --   & --    & --   \\ \hline
0.0 & 15 & 3.0 & 1.5 & 10 & 0.35 & 35  & 0.320 & 0.47 & 0.332 & 1.24 \\ 
0.2 & 15 & 3.0 & 1.5 & 10 & 0.35 & 35  & 0.348 & 0.95 & 0.368 & 1.40 \\ 
0.4 & 15 & 3.0 & 1.5 & 10 & 0.35 & 35  & 0.433 & 0.31 & 0.453 & 1.40 \\ 
0.6 & 15 & 3.0 & 1.5 & 10 & 0.35 & 35  & --    & --   & --    & --   \\ 
0.8 & 15 & 3.0 & 1.5 & 10 & 0.35 & 35  & --    & --   & --    & --   \\ \hline
0.2 & 9  & 1.8 & 0.9 & 75 & 0.60 & 100 & 0.610 & 0.85 & 0.640 & 1.60 \\ 
0.4 & 9  & 1.8 & 0.9 & 75 & 0.60 & 100 & 0.660 & 0.40 & 0.686 & 1.52 \\ 
0.5 & 9  & 1.8 & 0.9 & 75 & 0.60 & 100 & 0.730 & 0.65 & 0.746 & 1.32 \\ 
\end{tabular}
\end{center}
\caption{List of the initial 
condition parameters used for the complete set of simulations conducted in 
this paper (columns 1 to 7) as well as other characteristics of the evolution 
and final outcome of the simulations. Velocities are given in km s$^{-1}$, 
masses in $M_{\odot}$, temperatures in K, and times in Myr except for 
$t_{\rm evol}$ which is given in units of the $t_{\rm ff} \sim 0.05$ Myr.}
\label{tab:clump_initial}
\end{table*}

\subsection{The two suites of collisions}

A $10 M_\odot$ clump has equilibrium radius $R_0 \simeq 0.22\,{\rm pc}$, 
effective isothermal sound speed $c_0 \simeq 0.35\,{\rm km}\,{\rm s}^{-1}$ 
(corresponding to a cosmic mixture of H$_2$ and He at 35 K), central density 
$\rho_{\rm c} \simeq 2.6 \times 10^{-20}\,{\rm g}\,{\rm cm}^{-3}$ and boundary 
density $\rho_{\rm b} \simeq 9.1 \times 10^{-21}\,{\rm g}\,{\rm cm}^{-3}$. 
With $10 M_\odot$ clumps we simulate collisions having all possible 
combinations of 
Mach Number ${\cal M} = 5,\,10,\,{\rm and}\,15$ (corresponding to 
$|{\bf v}_{\rm clump}| = 0.5,\,1.0\,{\rm and}\,1.5\,{\rm km}\,{\rm s}^{-1}$) 
and offset parameter $b = 0.0,\,0.2,\,0.4,\,0.6\,{\rm and}\,0.8$ (the 
parameters adopted for each of the simulations we performed are listed 
in the first seven columns of Table \ref{tab:clump_initial}). 
The main results of this suite of simulations are presented in Section 
\ref{sec:low-mass}.

A $75 M_\odot$ clump has $R_0 \simeq 0.59\,{\rm pc}$, $c_0 \simeq 
0.60\,{\rm km}\,{\rm s}^{-1}$ (corresponding to a cosmic mixture of H$_2$ and 
He at $T \simeq 100\,{\rm K}$), $\rho_{\rm c} \simeq 1.1 \times 
10^{-21}\,{\rm g}\,{\rm cm}^{-3}$
and $\rho_{\rm b} \simeq 3.8 \times 10^{-22}\,{\rm g}\,{\rm cm}^{-3}$. 
With $75 M_\odot$ clumps we simulate only collisions having ${\cal M} = 9$ and 
$b = 0.2,\, 0.4\, {\rm and}\, 0.5$. We have 
chosen to study only this limited set of parameters for the intermediate-mass 
clump collision case, as a more detailed parameter study has been reported in 
Bhattal {\it et al}. \shortcite{BhattalMNRAS1998}. Here we aim to verify 
with higher numerical resolution the validity of their conclusions on the 
phenomenology of clump collisions. In particular, we would like to investigate 
whether the existence of three distinct mechanisms of star formation, which 
they identified in their clump collision simulations with varying $b$, depends 
on the low numerical resolution they employed. We have therefore chosen to 
perform simulations with only three values of $b$, as according to Bhattal 
{\it et al}. \shortcite{BhattalMNRAS1998} collisions with $b=0.2$, 0.5 and 0.4 
produce protostars through rotational fragmentation, filament fragmentation 
and the combination of the previous two, respectively. In addition, we use 
this suite of simulations to investigate the effect of increasing clump mass 
on the outcome of clump collisions (by comparing the results of this suite of 
simulations with those of the low-mass clump collisions with ${\cal M}=10$ 
and corresponding values for $b$). The parameters adopted for 
each of the three intermediate-mass simulations we performed are also listed 
in the first seven columns of Table \ref{tab:clump_initial}. 
The main results of this suite of simulations are presented in Section 
\ref{sec:amar}.

\section{SPH with Particle Splitting}{\label{sec:sph}}

\subsection{Standard self-gravitating SPH}{\label{sec:SPH}}

Smoothed Particle Hydrodynamics (SPH) 
\cite{GingoldMNRAS1977,LucyAJ1977,MonaghanARAnA1992,MonaghanRPP2005}, 
is a Lagrangian method for numerical hydrodynamics that assumes no symmetries 
or imposed grids, and is therefore very effective for treating problems that 
involve complex 3-dimensional geometries. SPH represents the fluid with an 
ensemble of ${\cal N}$ discrete but extended 
particles. The particles are overlapping, so that all intensive quantities 
can be treated as continuous functions both in time and space by averaging 
over neighbouring particles. To implement this, a smoothing kernel with 
compact support is used. The smoothing kernel describes the strength and 
extent of a particle's influence. We use the polynomial M4 kernel 
\cite{MonaghanAnA1985}.

The SPH particles move with the fluid and all hydrodynamical properties are 
calculated at the particle positions. To evolve the ensemble of SPH particles, 
we use a system of three equations, namely the continuity equation, Euler's 
equation and a barotropic equation of state (Eqn. \ref{equa:statecool}). The 
SPH formulations of the first two equations, giving the density and 
acceleration of particle $j$, are
\begin{equation}
\label{equa:rhofinal}
\rho_j \; = \; \sum_i \left\{ \frac{m_i}{\bar{h}_{ij}^3} \, W \left( 
\frac{|{\bf r}_{ij}|}{\bar{h}_{ij}} \right) \right\},
\end{equation}
\noindent and 
\begin{eqnarray} \nonumber
\frac{d{\bf v}_j}{dt} & = & -\,\sum_{i} \left\{ \left[ 
\frac{m_i}{\bar{h}_{ij}^4} 
\left( \frac{P_i}{\rho_i^2} + \frac{P_j}{\rho_j^2} + \Pi_{ij} \right) 
W'\left( \frac{|{\bf r}_{ij}|}{\bar{h}_{ij}} \right) \right. \right. \\
 & & \hspace{2.00cm} \left. \left. \,+\, \frac{m_i}{|{\bf r}_{ij}|^2} 
W^*\left( \frac{|{\bf r}_{ij}|}{\bar{h}_{ij}} \right)
\right] \frac{{\bf r}_{ij}}{|{\bf r}_{ij}|} \right\}\,.
\label{equa:dvdtfinal}
\end{eqnarray}
Here ${\bf r}_{ij} = {\bf r}_j - {\bf r}_i$ and $\bar{h}_{ij} = 
0.5 (h_i + h_j)$. $W(s)$, $W'(s)$ and $W^{*}(s)$ are the kernel, its 
derivative ($W'(s) \equiv {\rm d}W/{\rm d}s$) and its volume integral (i.e. 
the mass 
fraction contained within $s$ from the centre of the particle). 

The first term on the right hand side of Eqn. \ref{equa:dvdtfinal} gives the 
contribution to the acceleration from hydrodynamic and artificial viscosity 
forces. Artificial viscosity is included to prevent particle interpenetration 
at shocks. We use the standard artificial viscosity \cite{MonaghanARAnA1992}
\begin{equation}
\label{equa:Piartvisc}
\Pi_{ij} \; = \; \left \{
\begin{array}{ll}
\frac{- \alpha \mu_{ij} \bar{c}_{ij} + \beta \mu_{ij}^{2}}
{\bar{\rho}_{ij}}, & ({\bf v}_{ij} \cdot {\bf r}_{ij}) < 0; \\
0, & ({\bf v}_{ij} \cdot {\bf r}_{ij}) > 0,
\end{array} \right .
\end{equation}
\noindent where
\begin{equation}
\label{equa:mu}
\mu_{ij} \, = \, \frac{({\bf v}_{ij} \cdot {\bf r}_{ij})
\bar{h}_{ij}}{|{\bf r}_{ij}|^{2} \, + \, 0.01 \bar{h}_{ij}^{2}} \,, 
\end{equation}
${\bf v}_{ij} = {\bf v}_j - {\bf v}_i$, $\;{\bar{\rho}_{ij}} = 0.5 (\rho_i + 
\rho_j)$ 
and ${\bar{c}_{ij}} = 0.5 (c_i + c_j)$ (the average isothermal sound speed). 
We have taken the artificial viscosity parameters to be $\alpha=\beta=1$. A 
test simulation of two flows colliding at Mach Number $\mathcal{M}$=10 has 
shown that our code reproduces the expected density and velocity jump 
conditions of the shock.

The second term on the right hand side of Eqn. \ref{equa:dvdtfinal} gives the 
gravitational contribution to the acceleration of particle $j$. By using the 
fact that the mass of a particle is kernel smoothed, and by invoking Gauss' 
gravitational theorem, close gravitational interactions (that would otherwise 
give unphysically large accelerations) are softened. It is implicit in Eqn. 
\ref{equa:dvdtfinal} that we use the same smoothing length to soften both 
gravitational and hydrodynamical forces.

Due to the compact support of the kernel, the summations in Eqn. 
\ref{equa:rhofinal} and the first term on the right hand side of Eqn. 
\ref{equa:dvdtfinal} are not over all particles, but over the small number 
${\cal N}_{\rm neib}$ of nearby particles for which $|{\bf r}_{ij}| < 2 
\bar{h}_{ij} \;\;(\equiv h_i+h_j)$. In three dimensions SPH gives good results 
with ${\cal N}_{\rm neib} \sim 50$, and accordingly the smoothing 
length $h_i$ for particle $i$ is adjusted at each timestep so that 
${\cal N}_{\rm neib} \sim 50$.

Calculating the gravitational accelerations by a direct summation over all 
particle pairs, as implied by the right hand side of Eqn. 
\ref{equa:dvdtfinal}, is an ${\cal O}({\cal N}^2)$ process, and therefore 
prohibitively expensive for large ${\cal N}$. We avoid this expense by using 
the Tree-Code Gravity algorithm \cite{BarnesNATURE1986,HernquistApJSS1989}, 
which scales as ${\cal O}({\cal N}
\log{\cal N})$. With this 
algorithm the computational domain is divided recursively into a hierarchy of 
cells within cells until, at the lowest level, each drawn cell contains 
either a single particle or no particle, from which we naturally store 
only tree-leaves containing a single particle each. Then we can calculate the 
gravitational interaction between particle $j$ and particles at large 
distances from it, by treating these distant particles as a single point mass 
at the centre of mass of the cell to which they belong.

For the time evolution of our code we use a second order Range-Kutta 
integration scheme and multiple particle timesteps. A detailed description 
of our standard code is given in Turner {\it et al}. 
\shortcite{TurnerMNRAS1995}.

\subsection{The Jeans condition}

The Jeans condition requires that the Jeans mass $M_{\rm Jeans}$ be resolved 
throughout the computational domain, at all times. In SPH, the minimum 
resolvable mass is estimated to be $M_{\rm resolved} \sim {\cal N}_{\rm neib} 
m_{\rm ptcl}$, where $m_{\rm ptcl}$ is the mass of a single SPH particle 
\cite{BateMNRAS1997}. In a three-dimensionally extended medium, the Jeans mass 
is given by $M_{\rm Jeans} \sim G^{-3/2} \rho^{-1/2} c^3$. Thus the Jeans 
condition, $M_{\rm resolved} \lesssim M_{\rm Jeans}$, reduces to a maximum 
density that can be resolved with SPH particles of mass $m_{\rm ptcl}\,$,
\begin{equation}
\label{equa:maxdens}
\rho_{\rm max} \;\sim\; \frac{c^6}{G^3 \left({\cal N}_{\rm neib} 
m_{\rm ptcl} \right)^2} \,, 
\end{equation}
or equivalently a maximum SPH particle mass that can be used to model gas 
having density $\rho$,
\begin{equation}
m_{\rm max} \sim \frac{c^3}{G^{3/2} \rho^{1/2} 
{\cal N}_{\rm neib}}.
\end{equation}

With the barotropic equation of state that we are using (Eqn. 
\ref{equa:statecool}), the combination $c^3 / \rho^{1/2}$ reaches its 
minimum when $\rho \sim \rho_1$, i.e. when the gas switches from being 
approximately isothermal ($c \simeq c_{\rm s}$) for $\rho_0 < \rho < \rho_1$ 
to approximately adiabatic for $\rho > \rho_1$. Hence, the maximum SPH 
particle mass becomes
\begin{equation}
\label{equa:jeans}
m_{\rm max} \;\sim\; \frac{c_{\rm s}^3}{G^{3/2} \rho_1^{1/2}{\cal N}_{\rm neib}} 
\;\sim\; 5 \times 10^{-5} M_\odot \,;
\end{equation}
where the final evaluation has been made by substituting $c_{\rm s} \sim 
0.2\,{\rm km}\,{\rm s}^{-1}$, $\rho_1 \sim 10^{-14}\,{\rm g}\,{\rm cm}^{-3}$, 
and ${\cal N}_{\rm neib} \sim 50$.

A second -- but as it turns out less severe -- constraint on the mass of an 
SPH particle can be obtained by considering the shock compressed layer which 
forms between two colliding clumps, and requiring that the fragments into 
which it breaks up are resolved. Whitworth {\it et al}. 
\shortcite{WhitworthAnA1994} have shown, on the basis of linear perturbation 
analysis, that these fragments should have mass 
\begin{eqnarray} \nonumber
\label{equa:fragmentmass}
M_{\rm frag} & \sim & \frac{c_{\rm s}^3}{\left( G^3 \rho_{\rm clump} 
{\cal M} \right)^{1/2}} \\ \nonumber
 & \sim & \left( \frac{c_{\rm s}}{c_0} \right)^3 \frac{M_0}{{\cal M}^{1/2}} \\
 & \sim & \frac{M_\odot^{3/4} M_0^{1/4}}{{\cal M}^{1/2}} \,, \\ \nonumber
\end{eqnarray}
where we have obtained the second expression by substituting $\rho_{\rm clump} 
\sim c_0^6 / G^3 M_0^2$ for the pre-collision density of a clump in 
hydrostatic equilibrium, and the third expression by substituting for $c_0$ 
from Eqn. \ref{equa:larson}. If we now require that $M_{\rm resolved} 
\lesssim M_{\rm frag}$, the maximum SPH particle mass becomes
\begin{eqnarray} \nonumber
m_{\rm max}' & \sim & \frac{M_\odot^{3/4} M_0^{1/4}}{{\cal N}_{\rm neib} 
{\cal M}^{1/2}} \\
 & \sim & 2 \times 10^{-2} M_{\odot} \left( \frac{M_0}{M_\odot} \right)^{1/4} 
{\cal M}^{-1/2}\,. \\ \nonumber
\end{eqnarray}
Unless we consider very high-velocity collisions between very low-mass clumps, 
$m'_{\rm max}$ is much larger, and therefore less restrictive, than 
$m_{\rm max}$.

It follows that, if the Jeans condition is to be satisfied, a standard SPH 
code cannot model a collision between two $10 M_\odot$ clumps with fewer than 
$\sim 400,000$ SPH particles, and a collision between two $75 M_\odot$ clumps 
requires $\sim 3,000,000$ SPH particles. Simulations with such particle 
numbers are routinely performed at large parallel supercomputers 
\cite{BateMNRAS2003,BateMNRAS2005,JappsenAnA2005}, but on the 
smaller serial machines available to us they are prohibitive. In order 
to circumvent these prohibitive computational requirements, we have applied 
Particle Splitting \cite{KitsionasMNRAS2002}.

\subsection{Particle Splitting}

In Particle Splitting, we replace individual particles (parent particles) with 
small groups of particles (families of child particles) either globally in 
a predefined sub-region of the computational domain where we anticipate the 
need for greater resolution ({\it Nested Splitting}) 
or conditionally according to some locally defined criterion ({\it On-the-Fly 
Splitting}). Each parent particle is replaced by thirteen child particles, 
having masses $m_{\rm child} = m_{\rm parent}/13\,$ and smoothing lengths 
$h_{\rm child} = h_{\rm parent}/13^{1/3}$. One of these child particles is 
placed at the same position as its parent, and the other twelve child 
particles are positioned on the vertices of an hexagonal closed packed array, 
at equal distances $\ell = 1.5 h_{\rm child}$ from the central child particle, 
and from each other. The family of thirteen child particles is then rotated to 
an arbitrary orientation. By positioning the child particles in this way, we 
ensure (i) that the family of child particles is as close as possible to the 
spherically symmetric mass distribution of the parent particle, and (ii) that 
the local transient fluctuations due to Particle Splitting are minimised 
\cite{KitsionasMNRAS2002}.

Each child particle is given a velocity ${\bf v}_{\rm child}$ interpolated 
from the velocity field of its parent's neighbours $i$ (including the parent 
$j$ itself). This is formally given by
\begin{equation}
\label{equa:alpha1}
{\bf v}({\bf r}_{\rm child}) \; = \; \sum_{i,j} \left\{ 
\frac{m_i {\bf v}_i}{\rho_i \bar{h}_{ij}^3} \, W \left( 
\frac{|{\bf r}_{\rm child} - {\bf r}_{i}|}{\bar{h}_{ij}} \right) \right\}.
\end{equation}
The initial densities and accelerations of the child particles are then 
calculated using the standard SPH procedures (Eqns. \ref{equa:rhofinal} \& 
\ref{equa:dvdtfinal}). Subsequently, the child particles are also evolved 
with standard SPH procedures. The only difference is that, to mitigate 
interactions between adjacent particles having different masses, we have 
modified the scheme by which we calculate the smoothing lengths of particles. 
We now evolve $h_i$, the smoothing length for particle $i$, so that the 
radius $2 h_i$ contains $\sim 50$ times the mass of particle $i$, i.e. 
$50\,m_i$, rather than 50 other particles. This method can potentially 
introduce sampling errors in the calculation of the hydrodynamics quantities 
(see Eqn. \ref{equa:rhofinal} \& \ref{equa:dvdtfinal}), in case only a handful 
of neighbouring particles end 
up within $2 h_i$, e.g. a child particle surrounded only by more massive 
particles. We have not addressed such special cases in our current code as 
corresponding tests, presented in Kitsionas \shortcite{KitsionasPhD2000}, have 
shown that such errors are negligible. However, we caution the reader that 
this can be a potential problem with the implementation of Particle Splitting 
in other SPH codes.

In Kitsionas \& Whitworth \shortcite{KitsionasMNRAS2002} we have tested 
On-the-Fly Splitting on the standard Boss \& Bodenheimer 
\shortcite{BossApJ1979} problem. On-the-Fly Particle Splitting is invoked in 
response to the imminent violation of the Jeans condition, i.e. whenever the 
density is about to exceed $\rho_{\rm max}$ (Eqn. \ref{equa:maxdens}). SPH 
with Particle Splitting (either Nested or On-the-Fly) reproduces faithfully 
the results obtained with a standard high-resolution SPH simulation (using 
sufficient particles to satisfy the Jeans condition without Particle 
Splitting, but also using much more memory and CPU), {\em and} the results of 
high-resolution Adaptive-Mesh-Refinement Finite-Difference simulations 
\cite{TrueloveApJ1997,KleinTOKYO1998}. It also conforms to the analytic 
predictions of Inutsuka \& Miyama \shortcite{InutsukaApJ1992}.

In principle, Particle Splitting can be applied repeatedly to produce 
successive generations of ever smaller SPH particles, and hence ever finer 
mass-resolution. However, in the simulations presented here only one 
generation of child particles is needed. By invoking a single generation of 
Particle Splitting, we are able to start the simulations with SPH particles 
having $m_{\rm ptcl} = 13 m_{\rm max} \simeq 6.5 \times 10^{-4} M_\odot$ 
(Eqn. \ref{equa:jeans}). 
Therefore, initially a $10 M_\odot$ clump can be modelled with just 15,000 
particles, and a $75 M_\odot$ clump with 110,000 particles. The critical 
density at which these initial particles have to be split to avoid violating 
the Jeans condition is \begin{equation}
\rho_{\rm split} \;=\; 13^{-2}\,\rho_1 \;\simeq\; 6 \times 
10^{-17}\,{\rm g}\,{\rm cm}^{-3}\,.
\end{equation}
After this generation of Particle Splitting, adiabatic heating switches on 
before the simulation reaches its resolution limit again, and therefore, the 
Jeans condition is obeyed all the way up to the highest densities that can be 
achieved\footnote{The Jeans mass increases with increasing density after 
adiabatic heating switches on, as $M_{\rm Jeans} \propto c_{\rm s}^3 
\rho^{-1/2}$ and in this regime $c_{\rm s} \propto \rho^{1/3}$ ({\it cf}. 
Eqn. \ref{equa:statecool}). We note that the Jeans mass increases with 
increasing density for all adiabatic exponents greater than 4/3.}.

\begin{figure}

\setlength{\unitlength}{1mm}
\begin{picture}(80,80)
\includegraphics{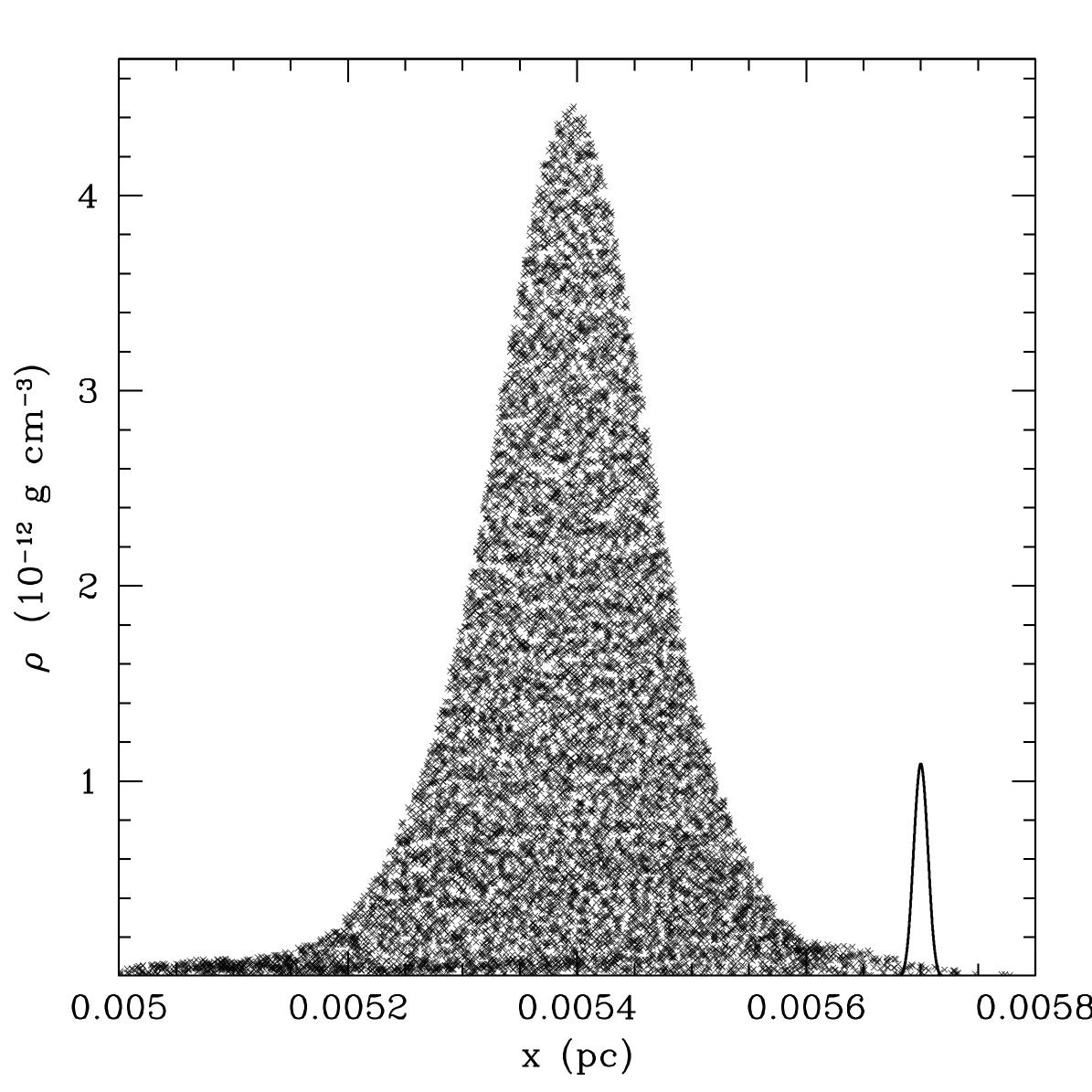}
\end{picture}

\caption{Linear density profile (particle plot) of one of the protostars 
formed at the end of the simulation with $M_0 = 10 M_\odot$, $b = 0.2$, 
${\cal M} = 10$, specifically the protostar at the bottom right of both 
panels of Fig. \ref{fig:b2m10xy}, with diameter of 
$\sim 7.5 \times 10^{-4}$ pc and density contrast of $\sim 3$ orders of 
magnitude. For comparison, the solid line illustrates the density profile of a 
single isolated particle with mass and smoothing length corresponding to those 
of particles close to the peak density of the protostar.}
\label{fig:profile}
\end{figure}

\subsection{Plots and diagnostics}{\label{sec:plots}}

We use grey-scale column density plots to present our results. All structure 
formed is contained within a single layer and therefore such plots are not 
greatly confused by projection effects. Column density plots are preferred to 
particle plots as the former give a more accurate representation of the total 
density field, and of what would be seen in optically thin molecular-line (or 
dust-continuum) radiation, assuming a uniform excitation temperature (or dust 
temperature). The figure captions give the linear size of each plot, the 
viewing axis, the time, and the range of the grey-scale (which is in all 
cases logarithmic, in units of g cm$^{-2}$).

In the sequel, we refer to all collapsing fragments identified in our 
simulations as protostars and to their discs as protostellar discs. We would 
like to note that the terms ``protostar'' and ``protostellar'' do 
not refer to pre-main-sequence (PMS) objects, as our simulations stop 
at much lower
density than that of PMS stars. 
Nevertheless, we use these terms for the fragments formed in our simulations, 
as all such fragments appear to have the characteristics of 
prestellar cores, i.e. 
collapsing cores not associated with $IR$ 
sources.

When a protostar forms, its linear density profile appears like a normal 
distribution around its peak density. At the end of our simulations, this 
profile is rather steep (e.g. Fig. \ref{fig:profile}). To infer its 
mass, we take the protostar to extend out to $d_{i} = 3\sigma_{i}$, where 
$\sigma_{i}$ is the FWHM in each of the three Cartesian axes $i = x, y, z$, 
and we sum the masses of the SPH particles that lie within all three diameters 
$d_{i}$. The radius of the protostar is given by the largest of the three 
radii $r_{i} = d_{i}/2$. This method gives good results for the 
mass and the radius of spherical as well as disc-like protostars. In essence, 
our method is similar to the Stutzki \& G{\"u}sten \shortcite{StutzkiApJ1990} 
clump-finding technique. Nevertheless, because of the low number of protostars 
formed in our simulations, we need not subtract each protostar from the 
density field before searching for the next one in the residuals. Therefore, 
we apply our method simultaneously to all protostars that are identified by 
eye in our simulations\footnote{A protostar is identified for the first 
time when its peak density exceeds $\sim$100 times the density of its 
surroundings. We consider this time of first identification as the formation 
time of the protostar.}. Moreover, we note that, 
when applied to a few of our simulations, the Williams {\it et al}. 
\shortcite{WilliamsApJ1994} clump-finding algorithm, as implemented 
for SPH by Klessen \& Burkert \shortcite{KlessenApJSS2000}, gives 
results very similar to those of our method. In Table \ref{tab:clump_initial}, 
we list the total mass of fragments, $M_{\rm frag}$, at the end of each 
simulation.

Because our simulations evolve at very high resolution and due to the 
Lagrangian nature of SPH, the timestep becomes extremely small in the densest 
parts of the computational domain, i.e. within the protostellar fragments 
formed. As a result, after a few thousand timesteps our simulations approach 
a state of suspended animation, when the very small particle distances within 
the fragments dominate the system evolution. We have arbitrarily chosen to 
terminate all simulations after 10,000 timesteps as it becomes extremely 
inefficient computationally to continue any further. To demonstrate that this 
is a reasonable choice, we have continued one of the simulations for another 
10,000 timesteps, during which the simulation advanced in physical time by 
only an additional $\sim 1\%$.

\begin{table*}
\begin{center}
\begin{tabular}{lp{30mm}p{43mm}p{42mm}p{33mm}}
${\cal M} \; \backslash \; b$ & 0.0 & 0.2 & 0.4 & 0.6 \\ \hline
5 & One spherical rotating object. $0.33 M_{\odot}$. No filaments. & One disc-like object. $0.7 M_{\odot}$. Spiral arms. No companions. A single filament. & Two disc-like objects. $1.11 M_{\odot}$ in total. Only most massive with spiral arms. Possible companions. Single filament. & Two well-separated rotating objects. $0.68 M_{\odot}$ in total. Only most massive with spiral arms. Possible companions. No filaments. \\ \hline 
10 & One spherical rotating object. $0.35 M_{\odot}$. Two filaments. & Two disc-like objects (+ a third forming). $1.0 M_{\odot}$ in total. Both spiral arms. No companions. Network of filaments. & Single disc-like object (+ a second forming). $0.48 M_{\odot}$. Spiral arms. Possible companions. Single filament. & No shock.\\ \hline 
15 & One disc-like rotating object. $0.47 M_{\odot}$. No spiral arms. Network of filaments. & Two disc-like objects (+ a third forming). $0.95 M_{\odot}$ in total. Only most massive with spiral arms. Possible companions. Well-defined network of filaments. & Single disc-like object. $0.31 M_{\odot}$. Spiral arms. No companions. No filaments. & No shock.\\ \hline 
\end{tabular}
\end{center}
\caption{Summary of simulations and most important results for the 
low-mass clump collisions ($M_{0}=10 M_{\odot}$) for the different values of 
$b$ and $\mathcal{M}$. Shocks do not form in any of the $b=0.8$ runs.}
\label{tab:lowclump}
\end{table*}

\section{$10 M_\odot$ clump collisions}{\label{sec:low-mass}}

In these simulations the colliding clumps each have mass $M_0 = 10 M_\odot$, 
radius $R_0 \simeq 0.22\,{\rm pc}$, and pre-collision effective sound speed 
$c_0 \simeq 0.35\,{\rm km}\,{\rm s}^{-1}$. In the centre-of-mass frame, 
their velocities are ${\bf v}_{\rm clump}$ and $-\,{\bf v}_{\rm clump}$ and 
they collide with impact parameter $bR_0$. We simulate collisions with 
Mach Numbers ${\cal M} = 5,\,10,\,15$ (corresponding to 
$|{\bf v}_{\rm clump}| \simeq 0.5,\,1.0,\,1.5\,{\rm km}\,{\rm s}^{-1}$), and 
$b = 0,\,0.2,\,0.4,\,0.6,\,0.8$ (corresponding to impact parameters 
$0,\,0.044,\,0.088,\,0.132,\,0.176\,{\rm pc}$).

Table \ref{tab:lowclump} summarises the main results of all low-mass 
collision simulations performed. In the following subsections we describe in 
more detail a subset of the simulations. First we hold $b = 0.2$ (constant) 
and increase ${\cal M}$ through ${\cal M} =$ 5 (\S\ref{sec:b2m5}), 
${\cal M} =$ 10 (\S\ref{sec:b2m10}), and ${\cal M} =$ 15 (\S\ref{sec:b2m15}); 
then we hold ${\cal M} = 5$ (constant) and increase $b$ through $b =$ 0.2 
(\S\ref{sec:b2m5}), $b =$ 0.4 (\S\ref{sec:b4m5}), and $b =$ 0.6 
(\S\ref{sec:b6m5}). In the following subsection we give a short overview 
of the expected outcome of clump collisions as discussed in Whitworth 
{\it et al}. \shortcite{WhitworthAnA1994,WhitworthMNRAS1994}.

\subsection{Analysis}

The outcome of a collision between clumps can be analysed in terms of three 
factors, namely (i) the mass, extent and lifetime of the shock compressed 
layer, (ii) the fragmentation scale in the shock compressed layer, and (iii) 
the net angular momentum of the shock compressed layer.

(i) {\it The mass, extent and lifetime of the shock compressed layer.} For 
a given clump mass $M_0$, the amount of mass which is shock compressed depends 
on the impact parameter. Evidently it is a maximum for head-on collisions 
($b = 0$), and decreases monotonically with increasing $b$, 
becoming negligible for $b \gtrsim 0.6$. For larger impact parameter, two 
further factors come into play. Firstly, the mass of the shock compressed 
layer is increased somewhat at low Mach Number because the clump trajectories 
are focused by their mutual gravitational attraction. The escape speed from 
the surface of a clump, $v_{\rm esc} = (2 G M_0 / R_0)^{1/2}$, is 
$\sim 0.6\,{\rm km}\,{\rm s}^{-1}$ for a $10 M_\odot$ clump, and so focusing 
is only important for the collisions with $|{\bf v}_{\rm clump}| \sim 
0.5\,{\rm km}\,{\rm s}^{-1}$ (i.e. ${\cal M} = 5$). Secondly, 
at high impact parameter and high Mach Number, a shock compressed layer 
forms but is then torn apart by shear before it can fragment.

(ii) {\it The fragmentation scale in the shock compressed layer.} As discussed 
in Whitworth {\it et al}. \shortcite{WhitworthAnA1994,WhitworthMNRAS1994}, the 
preferred fragment mass in a shock compressed layer is given by Eqn. 
\ref{equa:fragmentmass}, and the fragmentation scale (i.e. the mean 
separation between fragments, in the plane of the shock compressed layer) is 
\begin{eqnarray} \nonumber
\label{equa:fragmentsep}
L_{\rm frag} & \sim & \frac{c_{\rm s}}{\left( G \rho_{\rm clump} {\cal M} 
\right)^{1/2}} \\ \nonumber
 & \sim & \left( \frac{c_{\rm s}}{c_0} \right) \frac{R_0}{{\cal M}^{1/2}} \\
 & \sim & \left( \frac{M_0}{M_\odot} \right)^{-1/4} \frac{R_0}
{{\cal M}^{1/2}} \,. \\ \nonumber
\end{eqnarray}
We have obtained the second expression by substituting $\rho_{\rm clump} 
\sim c_0^6 / G^3 M_0^2$ for the pre-collision density of a clump in 
hydrostatic equilibrium and $R_0 \sim G M_0 / c_0^2$ for its radius, and the 
third expression by substituting for $c_0$ from Eqn. \ref{equa:larson}. It 
follows that for given clump mass $M_0$ and given (low) offset parameter $b$, 
increasing the Mach Number of the collision will decrease $L_{\rm frag}$, and 
hence increase the number of filaments into which the shock compressed layer 
fragments, and the number of protostars. At large $b$, the fragmentation scale 
is increased somewhat by the shear in the shock compressed layer. However, 
this effect is only significant for $b \gtrsim 0.5$, and the mass of the shock 
compressed layer is then negligible.

We should stress that the fragmentation of a shock compressed layer is 
essentially a two-dimensional process, initially. The fragmentation scale 
$L_{\rm frag}$ defines the size of proto-condensations in the plane of the 
layer. The thickness of the layer is much smaller than $L_{\rm frag}$. The 
number of filaments and protostars is therefore determined by the ratio of the 
lateral extent of the shock compressed layer to the fragmentation scale.

(iii) {\it The net angular momentum of the shock compressed layer.} The net 
angular momentum of the shock compressed layer increases with increasing 
impact parameter and increasing collision velocity, i.e. increasing $b$ and 
${\cal M}$. Increased angular momentum means that the ensemble of protostars 
has more collective orbital angular momentum, and so mergers between 
protostars are less likely. Additionally, it means that the material which 
accretes onto the individual protostellar discs from the filaments has higher 
specific angular momentum, and therefore spins them up more rapidly.

In the following five subsections we describe five simulations which 
illustrate the above features of collisions between $10 M_\odot$ clumps. 

\begin{figure*}

\setlength{\unitlength}{1mm}
\begin{picture}(80,90)
\includegraphics{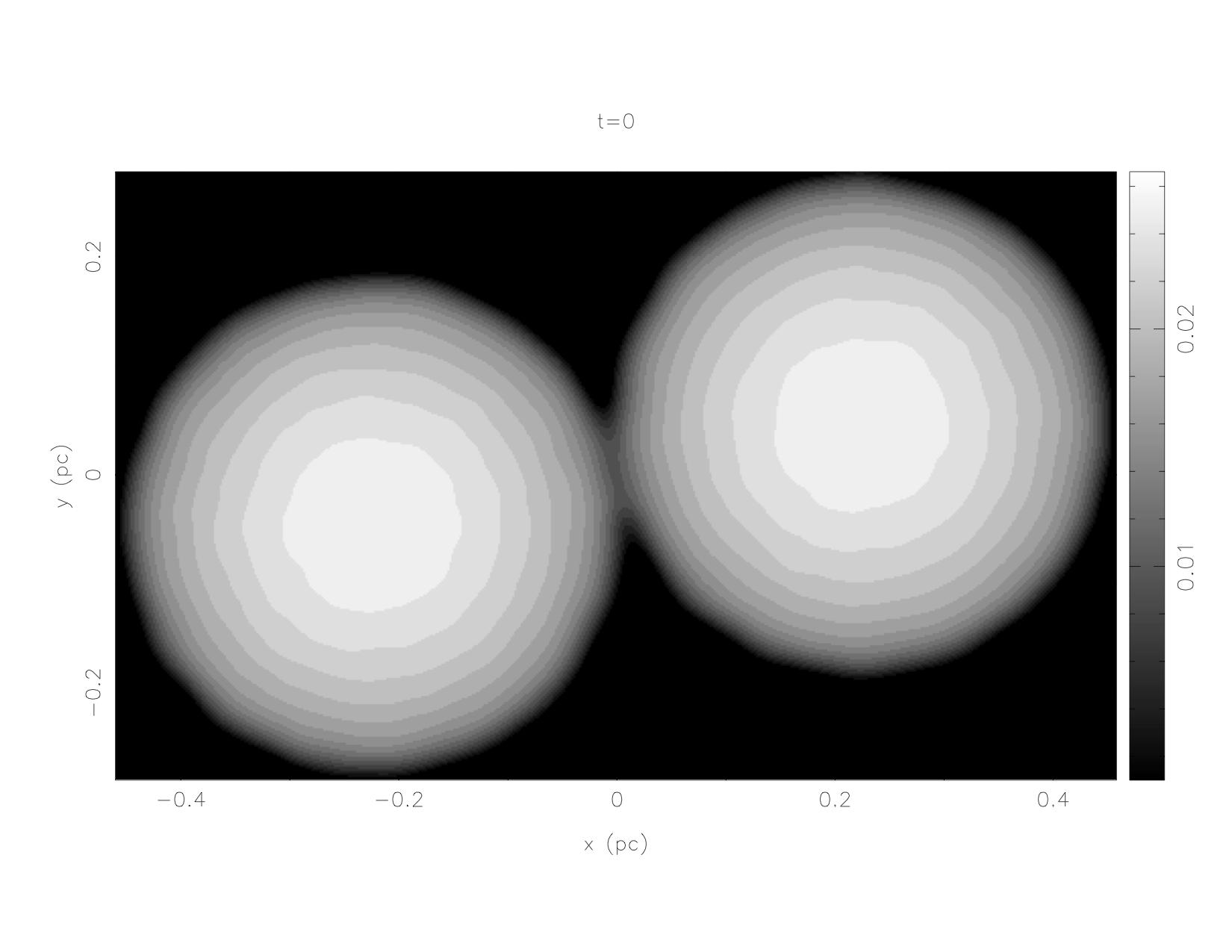}
\end{picture}
\begin{picture}(80,90)
\includegraphics{kitsionas_fig3b.ps}
\end{picture}

\caption{Column density plots for $M_0 = 10 M_\odot$, $b=0.2$, ${\cal M}=5$. 
{\it Left Panel.} Initial conditions viewed along the 
$z$-axis; $\Delta x = 0.92\,{\rm pc}$, $\Delta y = 0.56\,{\rm pc}$; the 
grey-scale is logarithmic, in units of g cm$^{-2}$, with sixteen equal 
intervals from $1.00 \times 10^{-3} \, {\rm g} \, {\rm cm}^{-2}$ to 
$2.69 \times 10^{-2} \, {\rm g} \, {\rm cm}^{-2}$ (or equivalently, adopting 
solar composition, $2.50 \times 10^{20} \, {\rm H}_2 \, {\rm cm}^{-2}$ to 
$6.73 \times 10^{21} \, {\rm H}_2 \, {\rm cm}^{-2}$). 
{\it Right Panel.} The end of the simulation ($t = 0.496\,{\rm Myr}$) viewed 
along the $z$-axis; $\Delta x = \Delta y = 0.016\,{\rm pc}$; sixteen-interval 
logarithmic grey-scale, in units of g cm$^{-2}$, from 
$2.40 \times 10^{-1} \, {\rm g} \, {\rm cm}^{-2}$ to 
$2.04 \times 10^{3} \, {\rm g} \, {\rm cm}^{-2}$ 
($6.00 \times 10^{22} \, {\rm H}_2 \, {\rm cm}^{-2}$ 
to $5.10 \times 10^{26} \, {\rm H}_2 \, {\rm cm}^{-2}$).}
\label{fig:b2m5xy}
\end{figure*}

\subsection{$M_0 = 10 M_\odot$, $b = 0.2$, ${\cal M} = 5$}{\label{sec:b2m5}}

The initial conditions are shown in the left panel of Fig. \ref{fig:b2m5xy}.
On-the-Fly Particle Splitting starts at $t_{\rm split}$ 
$\sim 0.463\,{\rm Myr}$. A single tumbling filament forms at 
$\sim 0.479\,{\rm Myr}$, and material from the filament accretes onto a single 
central primary protostar\footnote{We refer to protostars which form directly 
from the fragmentation of filaments as primary protostars.}. The filament is 
tumbling because the shock compressed layer from which it forms is tumbling, 
due to the orbital angular momentum in a clump-clump collision at finite 
impact parameter. The steadily increasing specific angular momentum of the 
material accreting onto the protostar from the filament creates an accretion 
disc around the protostar, and spiral arms develop in this disc. We are unable 
to follow the simulation for long enough to determine whether these arms 
become sufficiently self-gravitating to condense out as secondary companions.

At the end of the simulation ($\sim 0.496\,{\rm Myr}$; right 
panel of Fig. \ref{fig:b2m5xy}), the mass of the protostar is 
$\sim 0.70 M_{\odot}$ and its radius is $\sim 90\,{\rm AU}$. Its central 
density exceeds $\rho_1$, which implies that it has started to heat up 
adiabatically. The number of active particles has increased from 30,000 to 
55,300. 

\begin{figure*}

\setlength{\unitlength}{1mm}
\begin{picture}(50,90)
\includegraphics{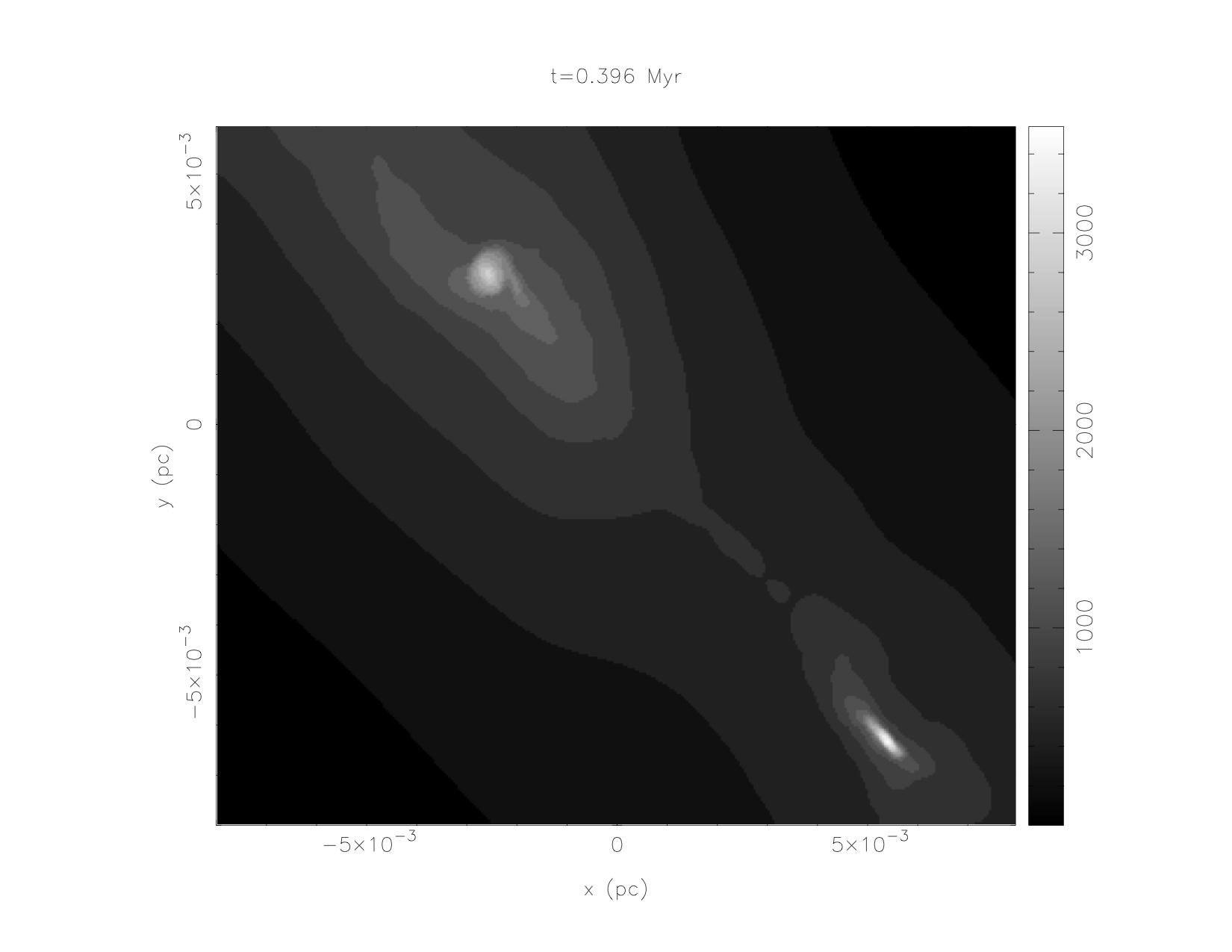}
\end{picture}
\begin{picture}(80,90)
\includegraphics{kitsionas_fig4b.ps}
\end{picture}

\caption{$M_0 = 10 M_\odot$, $b = 0.2$ ${\cal M} = 10$ at the end of the 
simulation ($t \sim 0.396\,{\rm Myr}$). 
{\it Left Panel.} View along the $z$-axis; $\Delta x = 0.016\,{\rm pc}, \; 
\Delta y = 0.014\,{\rm pc}$; sixteen-interval logarithmic grey-scale, 
in units of g cm$^{-2}$, from 
$2.14 \times 10^{-1} \, {\rm g} \, {\rm cm}^{-2}$ to 
$3.55 \times 10^{3} \, {\rm g} \, {\rm cm}^{-2}$ 
($5.35 \times 10^{22} \, {\rm H}_2 \, {\rm cm}^{-2}$ to 
$8.88 \times 10^{26} \, {\rm H}_2 \, {\rm cm}^{-2}$). 
{\it Right Panel.} View along the $y$-axis; $\Delta x = \Delta z = 
0.02\,{\rm pc}$; sixteen-interval logarithmic grey-scale, in units of 
g cm$^{-2}$, from 
$1.05 \times 10^{-1} \, {\rm g} \, {\rm cm}^{-2}$ to 
$1.78 \times 10^{3} \, {\rm g} \, {\rm cm}^{-2}$ 
($2.63 \times 10^{22} \, {\rm H}_2 \, {\rm cm}^{-2}$ to 
$4.45 \times 10^{26} \, {\rm H}_2 \, {\rm cm}^{-2}$).}
\label{fig:b2m10xy}
\end{figure*}

\subsection{$M_0 = 10 M_\odot$, $b = 0.2$, ${\cal M} = 10$}{\label{sec:b2m10}}

The initial conditions are again as in the left panel of Fig. 
\ref{fig:b2m5xy}. On-the-Fly Particle Splitting starts at $t_{\rm split} 
\sim 0.339\,{\rm Myr}$. By this stage, a network of tumbling filaments has 
started to form. At $\sim 0.360\,{\rm Myr}$, two protostars condense out of 
the filaments. The protostars are rapidly rotating, and accretion discs form 
around them shortly after their formation. The discs are approaching each 
other, and depending on their separation and their mutual alignment at 
periastron they are likely either to be captured into a binary, or to merge. 
There may be a third protostar starting to form a few timesteps before the end 
of the simulation.

At the end of the simulation ($\sim 0.396\,{\rm Myr}$; Fig. 
\ref{fig:b2m10xy}), the two protostellar discs have developed strong spiral 
arms, but they have not yet shown any inclination to fragment. The total mass 
of the two protostars is $\sim 1.0 M_{\odot}$. The more massive protostar has 
mass $0.59 M_{\odot}$ and radius $\sim 76\,{\rm AU}$. The less massive 
protostar has mass $0.41 M_{\odot}$ and radius $\sim 103\,{\rm AU}$. Their 
central densities exceed $\rho_1$, so they have started to heat up 
adiabatically. The minimum density in the filaments is $\rho_{\rm fil} 
\sim 2.8 \times 10^{-17} \,{\rm g}\,{\rm cm}^{-3}$ ($n_{{\rm H}_{2}} 
\sim 5 \times 10^{6} \,{\rm cm}^{-3}$). The number of active particles has 
increased from 30,000 to 64,300. 

\begin{figure*}
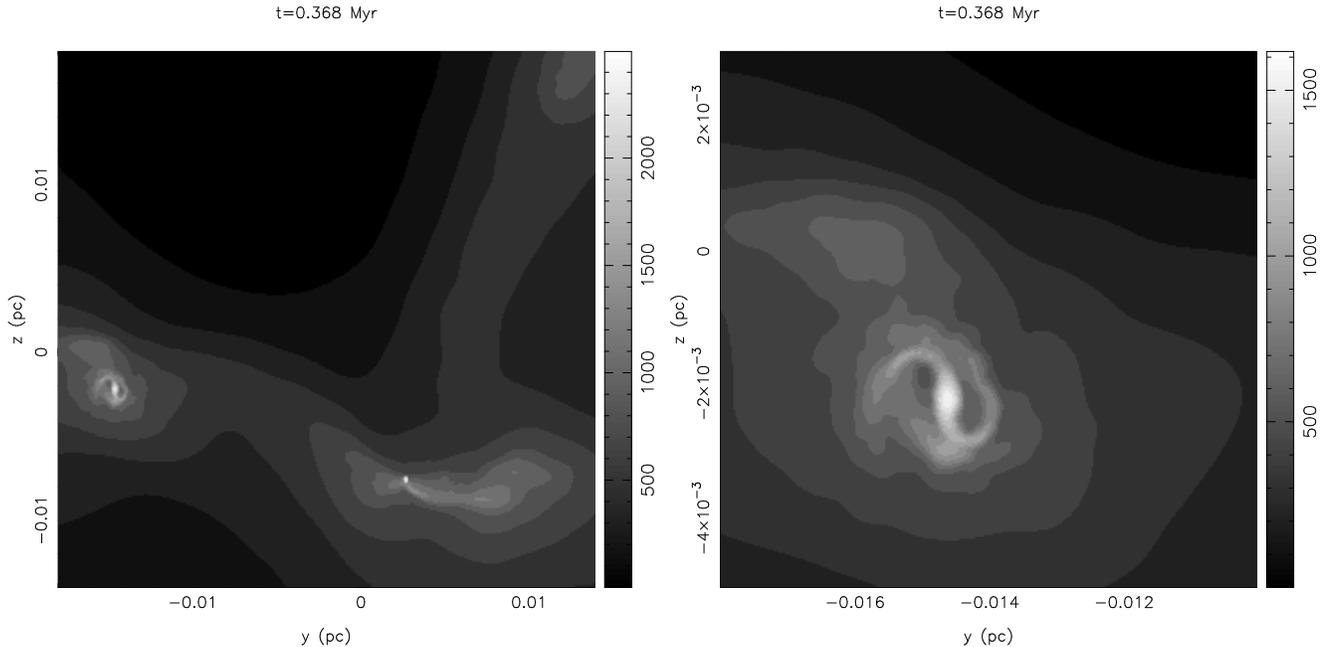


\setlength{\unitlength}{1mm}
\begin{picture}(80,90)
\includegraphics{kitsionas_fig5a.ps}
\end{picture}
\begin{picture}(80,90)
\includegraphics{kitsionas_fig5b.ps}
\end{picture}

\caption{$M_0 = 10 M_\odot$, $b = 0.2$, ${\cal M} = 15$ at the end of the 
simulation ($t = 0.368\,{\rm Myr}$), viewed along the $x$-axis. 
{\it Left Panel.} $\Delta y = \Delta z = 0.032\,{\rm pc}$; sixteen-interval 
logarithmic grey-scale, in units of 
g cm$^{-2}$, from 
$6.17 \times 10^{-2} \, {\rm g} \, {\rm cm}^{-2}$ to 
$2.51 \times 10^{3} \, {\rm g} \, {\rm cm}^{-2}$
($1.54 \times 10^{22} \, {\rm H}_2 \, {\rm cm}^{-2}$ to 
$6.28 \times 10^{26} \, {\rm H}_2 \, {\rm cm}^{-2}$). 
{\it Right Panel.} Zooming in on the protostar on the left edge of the left 
panel. $\Delta y = \Delta z = 0.008\,{\rm pc}$; sixteen-interval 
logarithmic grey-scale, in units of 
g cm$^{-2}$, from 
$1.51 \times 10^{-1} \, {\rm g} \, {\rm cm}^{-2}$ to 
$1.62 \times 10^{3} \, {\rm g} \, {\rm cm}^{-2}$ 
($3.78 \times 10^{22} \, {\rm H}_2 \, {\rm cm}^{-2}$ to 
$4.05 \times 10^{26} \, {\rm H}_2 \, {\rm cm}^{-2}$).}
\label{fig:b2m15yz}
\end{figure*}

\subsection{$M_0 = 10 M_\odot$, $b = 0.2$, ${\cal M} = 15$}{\label{sec:b2m15}}

Again, the initial conditions are as in the left panel of Fig. 
\ref{fig:b2m5xy}. A network of tumbling filaments forms at 
$\sim 0.306\,{\rm Myr}$. When compared with the filaments formed in the 
previous lower-velocity (${\cal M} = 10$) collision, the filaments of the 
${\cal M} = 15$ collision are better defined, and they form on the whole 
surface of the shock compressed layer, not just at its centre. On-the-Fly 
Particle Splitting starts at $t_{\rm split} \sim 0.336\,{\rm Myr}$. Two 
protostars condense out of the central filament at $\sim 0.348\,{\rm Myr}$. 
These protostars are rapidly rotating, and accretion discs form around them 
shortly after their formation. The disc of the more massive (primary) 
protostar develops strong spiral arms. There are density enhancements at the 
points where the accretion flows intercept the spiral arms (right panel of 
Fig. \ref{fig:b2m15yz}). These density enhancements may subsequently become 
self-gravitating and condense out as secondary companions to the primary. 
There may be a third object starting to condense out in one of the other 
filaments.

At the end of the simulation ($\sim 0.368\,{\rm Myr}$; Fig. 
\ref{fig:b2m15yz}), the total mass of the two protostars is 
$\sim 0.95 M_{\odot}$. There is also another $\sim 0.1 M_{\odot}$ associated 
with the spiral arms. The more massive protostar has mass $0.53 M_{\odot}$ 
and radius $\sim 42\,{\rm AU}$. The other protostar has mass $0.42 M_{\odot}$ 
and radius $\sim 35\,{\rm AU}$. Their central densities exceed $\rho_1$, so 
they have started to heat up adiabatically. The minimum density in the 
filaments is $\rho_{\rm fil} \sim 10^{-16}\,{\rm g} \,{\rm cm}^{-3}$ 
($n_{{\rm H}_{2}} \sim 10^{7}\,{\rm cm}^{-3}$). The number of active 
particles has increased from 30,000 to 70,200.

\begin{figure*}
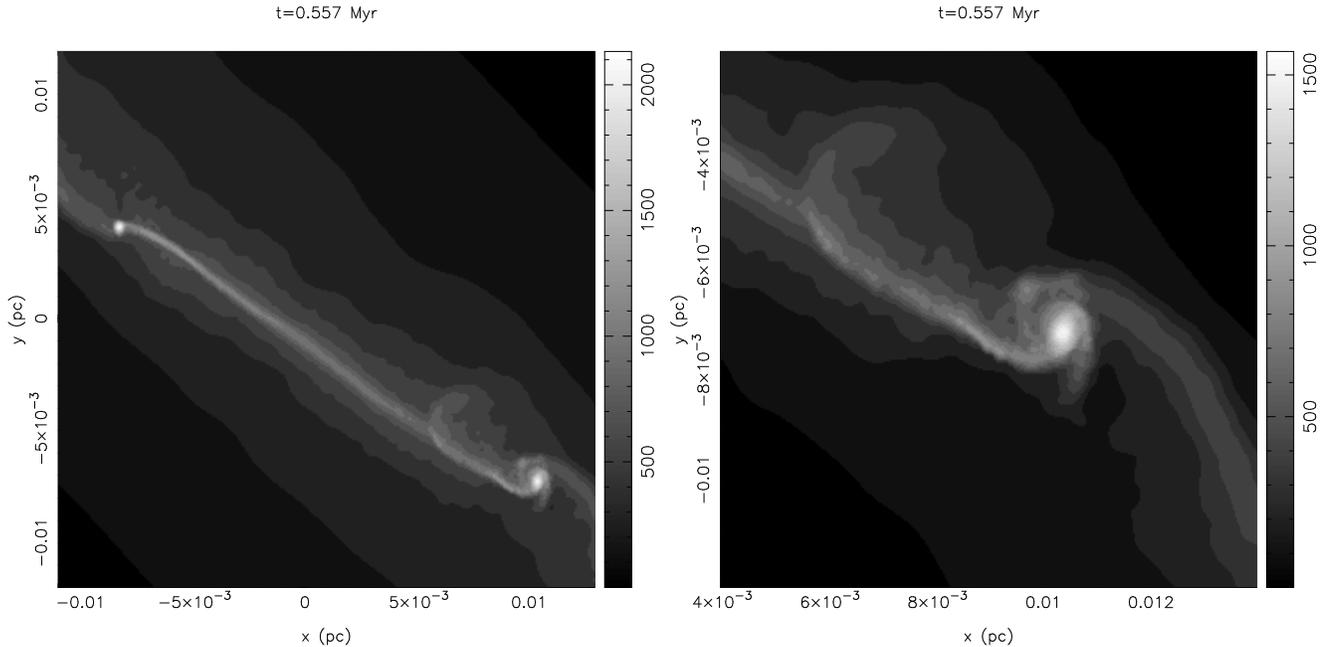


\setlength{\unitlength}{1mm}
\begin{picture}(80,90)
\includegraphics{kitsionas_fig6a.ps}
\end{picture}
\begin{picture}(80,90)
\includegraphics{kitsionas_fig6b.ps}
\end{picture}

\caption{$M_0 = 10 M_\odot$, $b = 0.4$, ${\cal M} = 5$ at the end of the 
simulation ($t = 0.557\,{\rm Myr}$), viewed along the $z$-axis. 
{\it Left Panel.} $\Delta x = \Delta y = 0.024\,{\rm pc}$; 
sixteen-interval logarithmic grey-scale, in units of 
g cm$^{-2}$, from 
$8.13 \times 10^{-2} \, {\rm g} \, {\rm cm}^{-2}$ to 
$2.14 \times 10^{3} \, {\rm g} \, {\rm cm}^{-2}$ 
($2.03 \times 10^{22} \, {\rm H}_2 \, {\rm cm}^{-2}$ to 
$5.35 \times 10^{26} \, {\rm H}_2 \, {\rm cm}^{-2}$). 
{\it Right Panel.} Zooming in on the protostar in the bottom right hand corner 
of the left panel. $\Delta x = \Delta y = 0.01\,{\rm pc}$.; 
sixteen-interval logarithmic grey-scale, in units of 
g cm$^{-2}$, from 
$2.24 \times 10^{-1} \, {\rm g} \, {\rm cm}^{-2}$ to 
$1.58 \times 10^{3} \, {\rm g} \, {\rm cm}^{-2}$ 
($5.60 \times 10^{22} \, {\rm H}_2 \, {\rm cm}^{-2}$ to 
$3.95 \times 10^{26} \, {\rm H}_2 \, {\rm cm}^{-2}$).}
\label{fig:b4m5xy}
\end{figure*}

\subsection{$M_0 = 10 M_\odot$, $b = 0.4$, ${\cal M} = 5$}{\label{sec:b4m5}}

On-the-Fly Particle Splitting starts at $t_{\rm split} \sim 0.525\,{\rm Myr}$. 
A single tumbling filament forms at $\sim$0.532 Myr. Two protostars form 
towards the two ends of the filament at $\sim$0.535 Myr. The protostars 
rotate rapidly, and accretion discs form around them shortly after their 
formation. One of the two protostellar discs develops strong spiral arms 
(right panel of Fig. \ref{fig:b4m5xy}), and there are density enhancements 
at the points where the spiral arms interact with the accretion flow. 
Secondary companions to the primary protostar may subsequently form from 
these enhancements. 

At the end of the simulation ($\sim 0.557\,{\rm Myr}$; Fig. \ref{fig:b4m5xy}), 
the total mass of the two protostars is $\sim 1.11 M_{\odot}$ (left panel of 
Fig. \ref{fig:b4m5xy}). The more massive protostar has mass $0.67 M_{\odot}$ 
and radius $\sim 77\,{\rm AU}$. The other protostar has mass $0.44 M_{\odot}$ 
and radius $\sim 52\,{\rm AU}$. Their central densities exceed $\rho_1$, so 
they have started to heat up adiabatically. The protostars are separated by 
$\sim 3500\,{\rm AU}$. The minimum density in the filaments is 
$\rho_{\rm fil} \sim 2.43 \times 10^{-17}\,{\rm g}\,{\rm cm}^{-3}$ 
($n_{{\rm H}_{2}} \sim 5 \times 10^{6}\,{\rm cm}^{-3}$). The number of active 
particles has increased from 30,000 to 79,400. 

\begin{figure}

\setlength{\unitlength}{1mm}
\begin{picture}(80,90)
\includegraphics{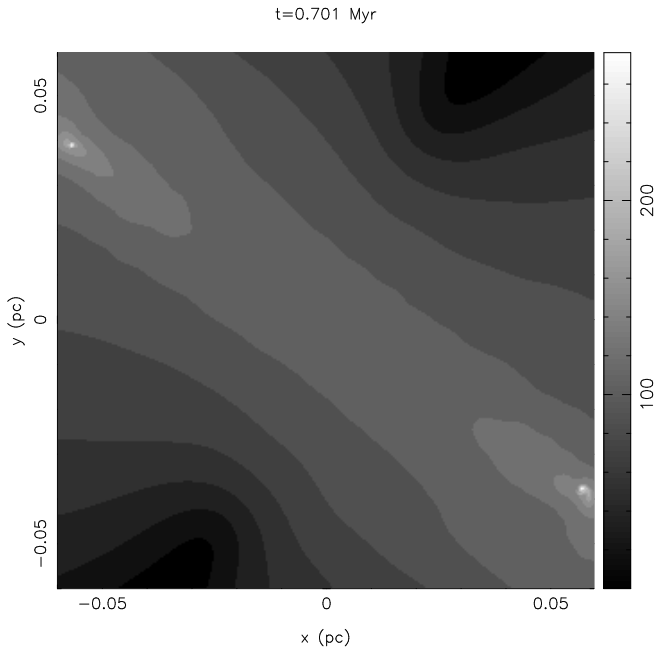}
\end{picture}

\caption{$M_0 = 10 M_\odot$, $b = 0.6$, ${\cal M} = 5$, at the end of 
the simulation ($t = 0.701\,{\rm Myr}$), viewed along the $z$-axis; 
$\Delta x = \Delta y = 0.12\,{\rm pc}$;
sixteen-interval logarithmic grey-scale, in units of 
g cm$^{-2}$, from
$2.45 \times 10^{-4} \, {\rm g} \, {\rm cm}^{-2}$ to 
$2.75 \times 10^{2} \, {\rm g} \, {\rm cm}^{-2}$ 
($6.13 \times 10^{19} \, {\rm H}_2 \, {\rm cm}^{-2}$ to 
$6.88 \times 10^{25} \, {\rm H}_2 \, {\rm cm}^{-2}$).}
\label{fig:b6m5xy}
\end{figure}

\subsection{$M_0 = 10 M_\odot$, $b = 0.6$, ${\cal M} = 5$}{\label{sec:b6m5}}

On-the-Fly Particle Splitting starts at $t_{\rm split} \sim 0.644\,{\rm Myr}$. 
The two clumps move a long way into each other before the density increases 
significantly. Two single well-separated protostars form at 
$\sim 0.678\,{\rm Myr}$, but no filaments are formed in this collision. 
The protostars rotate rapidly, and accretion discs form around them shortly 
after their formation. One of the two protostellar discs develops strong 
spiral arms (bottom right protostar in Fig. \ref{fig:b6m5xy}).

At the end of the simulation ($\sim 0.701\,{\rm Myr}$; Fig. \ref{fig:b6m5xy}), 
the total mass of the two protostars is $\sim 0.68 M_{\odot}$. The more 
massive protostar has mass $0.40 M_{\odot}$ and radius $\sim 53\,{\rm AU}$. 
The other protostar has mass $0.28 M_{\odot}$ and radius $\sim 85\,{\rm AU}$. 
Their central densities exceed $\rho_1$, and so they have started to heat up 
adiabatically. The separation between the two protostars is 
$> 30,000\,{\rm AU}$, and they are unbound. The number of active particles has 
increased from 30,000 to 50,000. 



\begin{figure}

\setlength{\unitlength}{1mm}
\begin{picture}(80,80)
\includegraphics{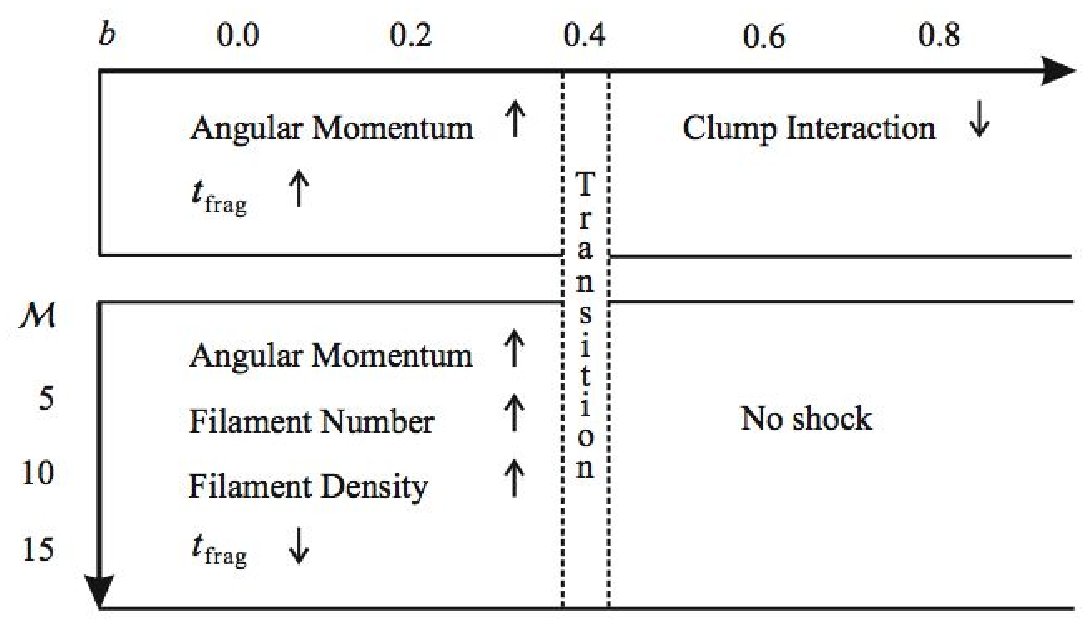}
\end{picture}

\caption{Dependence of different quantities and phenomena on the 
increasing values of $b$ and $\mathcal{M}$ for simulations of clump-clump 
collisions. Upward and downward pointing arrows indicate increasing and 
decreasing quantities, respectively. Note that the parameter space is 
divided in two sections: low-$b$ 
collisions produce stronger shocks. Large-$b$ collisions reduce the 
clump interaction. The transition happens at $b$=0.4.}
\label{fig:lowclump1}
\end{figure}

\subsection{Discussion of $10 M_\odot$ collisions}

As ${\cal M}$ is increased (with $M_0 = 10 M_\odot$ and $b = 0.2$ held 
constant), there are two main trends. Firstly, the fragmentation scale of the 
shock compressed layer, $L_{\rm frag}$, becomes smaller, and therefore the 
layer breaks up into more filaments and forms more protostars. Secondly, the 
ensemble of protostars thus produced has more orbital angular momentum, and 
therefore mergers are less likely.

As $b$ is increased (with $M_0 = 10 M_\odot$ and ${\cal M} = 5$ held 
constant), there are two main trends. Firstly, with larger impact parameter 
the mass of the shock compressed layer is reduced, and $L_{\rm frag}$ is 
slightly increased. Consequently, fewer protostars are formed -- and indeed 
for $b \gtrsim 0.6$ no protostars are formed. Secondly, the pattern of 
fragmentation changes, in the sense that for higher impact parameter i) two 
protostars form but with one being mainly material from one clump, and the 
other mainly material from the other clump; ii) the two protostars thus 
formed are not bound to one another. Fig. \ref{fig:lowclump1} summarises 
the above main conclusions.

Accretion rates onto the protostars are typically $1\;{\rm to}\;5 
\times 10^{-5} M_\odot \,{\rm yr}^{-1}$. There is no obvious dependence of the 
accretion rate on the collision parameters.

As we discussed in Section \ref{sec:eos}, the gas that gets compressed
in the shock at the collision interface cools radiatively down to 
$\sim 10\,{\rm K}$ and then remains isothermal for a few orders of 
magnitude in density before it gets heated adiabatically. Since gas 
fragmentation occurs in this isothermal regime, we consider the free-fall 
time, $t_{\rm ff}$, i.e. the most-commonly used measure of the fragmentation 
time-scale, to be that corresponding to gas of density 
$1\;{\rm to}\;2 \times 10^{-18}$ g cm$^{-3}$, i.e the density at which gas 
reaches the isothermal temperature of 10 K for the first time ({\it cf}. 
Fig. \ref{fig:eos}). This gives $t_{\rm ff} \sim 0.05$ Myr. Indeed, the time 
of fragmentation, $t_{\rm frag}$ (which is identical to the protostar 
formation time according to the definition given in Section \ref{sec:plots}), 
is in all simulations delayed by approximately one $t_{\rm ff}$ with respect 
to the time at which the gas in the simulation reached, for the first time, a 
density of $\sim 10^{-18}$ g cm$^{-3}$. This means that there are two main 
phases of evolution in the simulations: in the first phase the dense shock 
compressed layer is forming (while the gas is cooling) and in the second 
(isothermal) phase the gas in this layer fragments. 
In Table \ref{tab:clump_initial}, we list the fragmentation and the final 
time of each simulation, $t_{\rm frag}$ and $t_{\rm end}$ respectively (both 
given in Myr), as well as the approximate duration of the second of the two 
phases defined above, which we call $t_{\rm evol}$ (given in units of the 
$t_{\rm ff}$), for which we have concluded that all collisions fragment 
{\it exactly} one $t_{\rm ff} \sim 0.05$ Myr after the gas reached a 
temperature of 10 K for the first time.

As can be seen from Table \ref{tab:clump_initial}, the fragmentation time,
$t_{\rm frag}$, decreases with increasing Mach Number, ${\cal M}$, and 
decreasing offset parameter, $b$, as the formation of the shock compressed 
layer progresses faster with increasing clump velocity as well as in 
collisions where the clumps collide closer to head-on. 

In fact, the shock compressed layer breaks first into filaments which 
subsequently fragment to produce the (primary) protostars. Fragmentation of 
the filaments is, therefore, the main star formation mechanism at play in 
these low-mass clump collisions. It is the combination of gas cooling and 
self-gravity that are responsible for this behaviour. In particular, when we 
repeated the simulation with $b = 0.2$, ${\cal M} = 10$ without taking into
account the gas self-gravity, the two clumps quickly passed through each other 
as the shock compressed layer never became sufficiently dense (only by a 
factor of $\sim 2.5$ orders 
of magnitude) to break up into filaments and/or protostars. Repeating the same 
simulation, this time including the gas self-gravity but not allowing for gas 
cooling (i.e. the gas was isothermal at 35 K), no shock compressed layer 
formed but instead a single very massive ($\gtrsim 2.5 M_{\odot}$) and highly 
bound object formed, which was rotating very fast and was attended by a disc 
with spiral arms.

As can be seen from Tables \ref{tab:clump_initial} \& \ref{tab:lowclump}, 
our simulations end at a stage when only about 1-5\% of the total available 
gas mass has turned into protostars. Taking this value as a lower limit for 
the star formation efficiency (SFE) of the clump-clump collisions we model, 
we can also estimate an SFE upper limit from the gas mass that ends up
gravitationally bound in the filaments. This gives a value of $\sim$ 30-35\% 
(e.g. for the $b = 0.2$, ${\cal M} = 10$ run the bound mass in the filaments 
is $\sim 6.7 M_{\odot}$). The early termination of the simulations does not 
allow us to measure the SFE in a more precise way than these estimated 
upper and lower limits. One can also use the fact that the protostars formed 
here show high mass accretion rates, characteristic of Class 0 objects, and 
estimate an expected final mass for the protostars $\sim 1$ Myr after their 
formation (using mass accretion rates lower by one or two orders of magnitude 
for the duration of the Class I phase\footnote{We assume, very crudely, 
that the Class 0 phase lasts for $\sim 0.1$ Myr and the Class I phase 
for $\sim 1$ Myr, after which the mass accretion rate becomes negligible.}).
In this way one ends up with a final mass of $1-2 M_{\odot}$ per primary 
protostar formed. Using also the fact that 1-2 primary protostars form on 
average in our simulations\footnote{The number of Jeans masses in the 
filaments is on average of order 2 at the end of the simulations.}, we 
infer a SFE of order 10-20\% for the low-mass collisions. Based on the number 
of Jeans masses in the filaments, we expect that in the high ${\cal M}$, 
$b = 0.2$ collisions a third primary protostar will also form, as the number 
of Jeans masses in the filaments at the end of these collisions is of order 3. 
Thus, we expect the SFE of such collisions to be higher than the average, but 
still in the 10-20\% range.

In some of our simulations, spiral arms form in the protostellar discs and we 
find some evidence for interactions between the spiral arms and the accretion 
flows. However, our simulations stop at an early stage of the disc evolution 
due to timestep constraints. Thus, we can not confirm that such interactions 
are efficient in forming companions to the protostars. Formation of 
secondaries by accretion-induced rotational instabilities and/or disc-disc 
interactions would increase the SFE. We note, however, that the Toomre $q$ 
parameter\footnote{To calculate $q$, we use the average values of the 
sound speed, density, velocity, distance from the disc centre, and distance 
from the disc midplane for all particles in a disc.} for most of the discs 
(e.g. both discs of the $b = 0.2$, ${\cal M} = 10$ collision; see Fig. 
\ref{fig:b2m10xy}) is steadily decreasing and it is $q \gtrsim 1$ at the end 
of the simulations.

\begin{figure*}
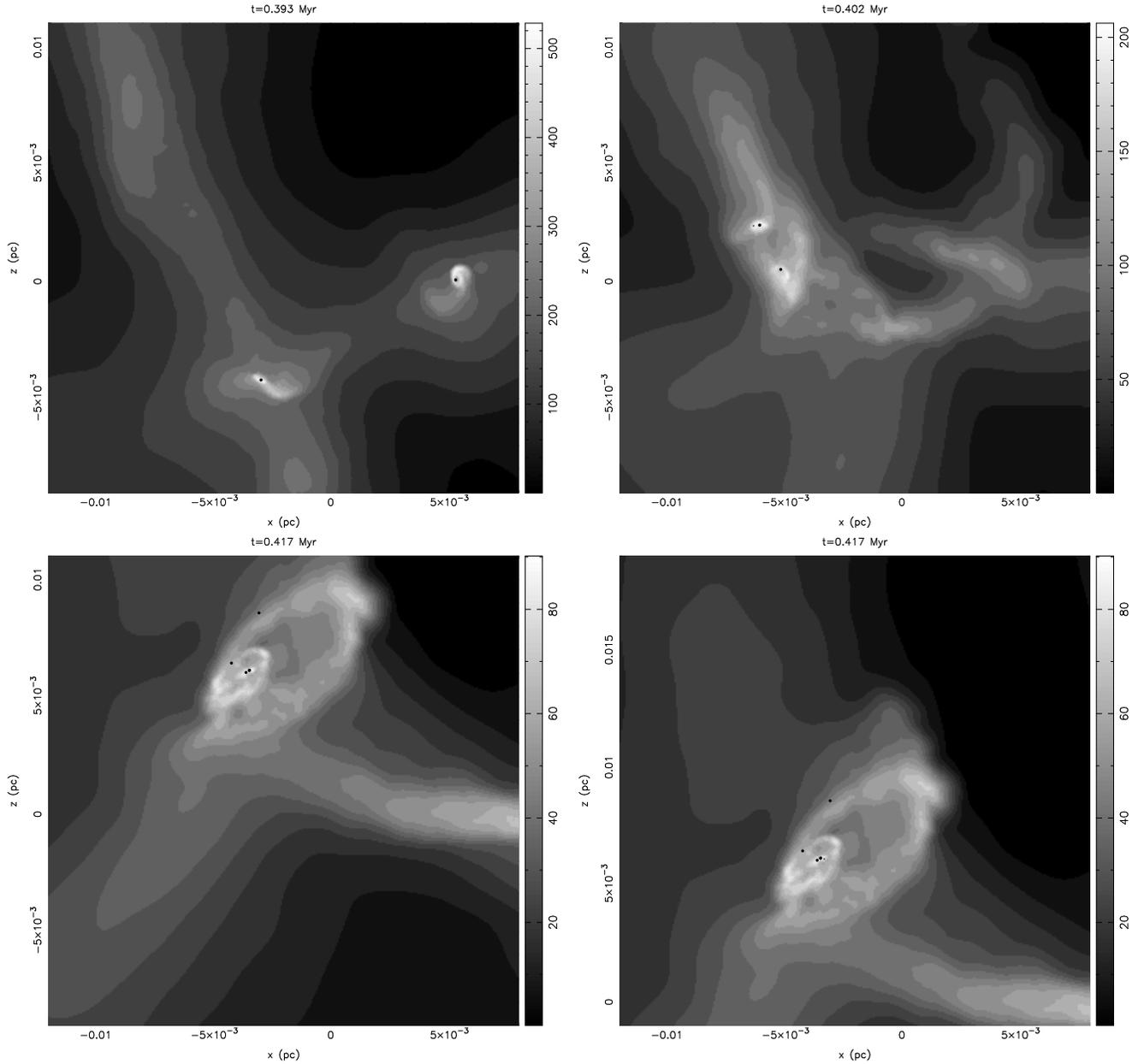


\setlength{\unitlength}{1mm}
\begin{picture}(80,85)
\includegraphics{kitsionas_fig9a.ps}
\end{picture}
\begin{picture}(80,85)
\includegraphics{kitsionas_fig9b.ps}
\end{picture}
\begin{picture}(80,85)
\includegraphics{kitsionas_fig9c.ps}
\end{picture}
\begin{picture}(80,85)
\includegraphics{kitsionas_fig9d.ps}
\end{picture}

\caption{Continuing the $M_0 = 10 M_\odot$, $b = 0.2$ ${\cal M} = 10$ 
simulation by using sink particles (overlayed solid circles; note that the 
symbol used for the sinks is larger than the actual sink radius at the scale 
of these plots). View along the $y$-axis; $\Delta x = \Delta z = 
0.02\,{\rm pc}$ in all panels. {\it Upper Left Panel.} 
$t \sim 0.393\,{\rm Myr}$, i.e. comparable to the right panel of Fig. 
\ref{fig:b2m10xy} (the end of the corresponding simulation {\it without} 
sinks); sixteen-interval logarithmic grey-scale, in units of 
g cm$^{-2}$, from 
$3.31 \times 10^{-1} \, {\rm g} \, {\rm cm}^{-2}$ to 
$5.25 \times 10^{2} \, {\rm g} \, {\rm cm}^{-2}$ 
($8.28 \times 10^{22} \, {\rm H}_2 \, {\rm cm}^{-2}$ to 
$1.31 \times 10^{26} \, {\rm H}_2 \, {\rm cm}^{-2}$). 
{\it Upper Right Panel.} $t \sim 0.402\,{\rm Myr}$; 
sixteen-interval logarithmic grey-scale, in units of 
g cm$^{-2}$, from 
$2.88 \times 10^{-1} \, {\rm g} \, {\rm cm}^{-2}$ to 
$2.04 \times 10^{2} \, {\rm g} \, {\rm cm}^{-2}$ 
($7.20 \times 10^{22} \, {\rm H}_2 \, {\rm cm}^{-2}$ to 
$5.10 \times 10^{25} \, {\rm H}_2 \, {\rm cm}^{-2}$).
{\it Both Lower Panels.} $t \sim 0.417\,{\rm Myr}$; the {\it Bottom Right 
Panel} is an exact copy of the {\it Bottom Left Panel} after having shifted 
the latter by 0.008 pc along the $z$-axis (in order to allow for direct 
comparisons with the corresponding plots of Fig. \ref{fig:new2}); 
sixteen-interval logarithmic grey-scale, in units of 
g cm$^{-2}$, from 
$2.45 \times 10^{-1} \, {\rm g} \, {\rm cm}^{-2}$ to 
$9.12 \times 10^{1} \, {\rm g} \, {\rm cm}^{-2}$ 
($6.13 \times 10^{22} \, {\rm H}_2 \, {\rm cm}^{-2}$ to 
$2.28 \times 10^{25} \, {\rm H}_2 \, {\rm cm}^{-2}$).}
\label{fig:new1}
\end{figure*}

\begin{figure*}
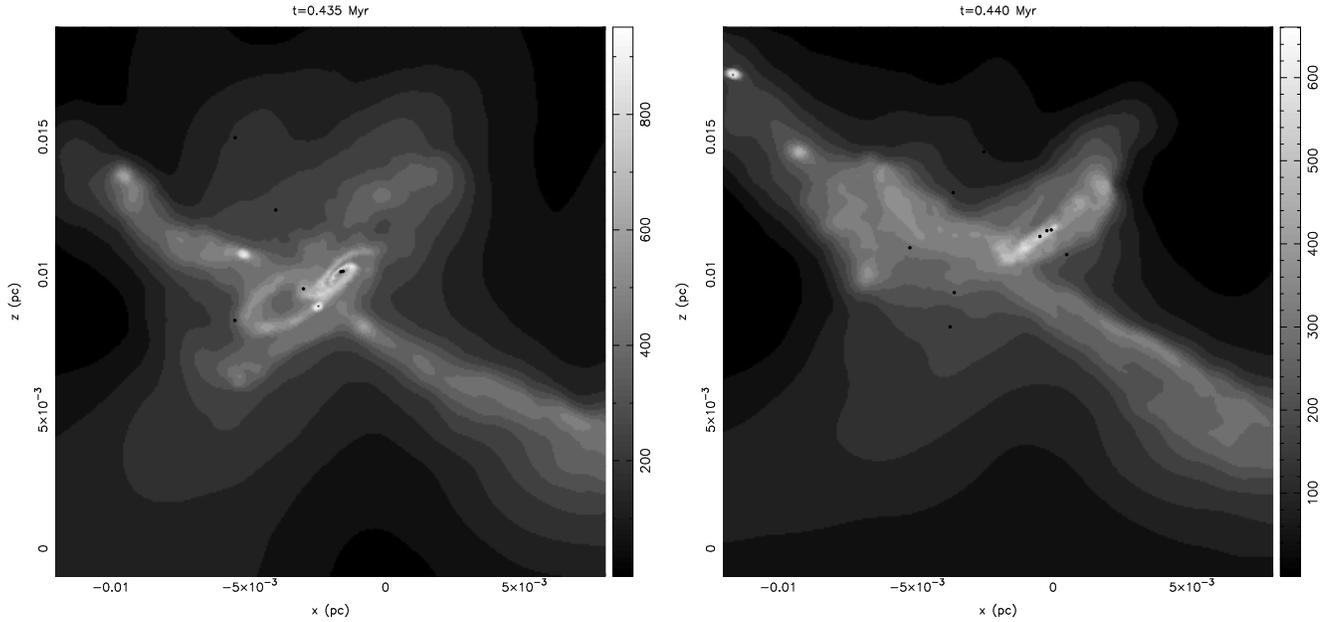


\setlength{\unitlength}{1mm}
\begin{picture}(80,85)
\includegraphics{kitsionas_fig10a.ps}
\end{picture}
\begin{picture}(80,85)
\includegraphics{kitsionas_fig10b.ps}
\end{picture}

\caption{Continuing the $M_0 = 10 M_\odot$, $b = 0.2$ ${\cal M} = 10$ 
simulation by using sink particles (overlayed solid circles; note that the 
symbol used for the sinks is larger than the actual sink radius at the scale 
of these plots). View along the $y$-axis; $\Delta x = \Delta z = 
0.02\,{\rm pc}$ in both panels (to be compared directly with the bottom right 
panel of Fig. \ref{fig:new1}). {\it Left Panel.} $t \sim 0.435\,{\rm Myr}$; 
sixteen-interval logarithmic grey-scale, in units of 
g cm$^{-2}$, from 
$3.09 \times 10^{-1} \, {\rm g} \, {\rm cm}^{-2}$ to 
$9.55 \times 10^{2} \, {\rm g} \, {\rm cm}^{-2}$ 
($7.73 \times 10^{22} \, {\rm H}_2 \, {\rm cm}^{-2}$ to 
$2.39 \times 10^{26} \, {\rm H}_2 \, {\rm cm}^{-2}$). 
{\it Right Panel.} At the end of the simulation ($t \sim 0.440\,{\rm Myr}$); 
sixteen-interval logarithmic grey-scale, in units of 
g cm$^{-2}$, from 
$3.39 \times 10^{-1} \, {\rm g} \, {\rm cm}^{-2}$ to 
$6.61 \times 10^{2} \, {\rm g} \, {\rm cm}^{-2}$ 
($8.48 \times 10^{22} \, {\rm H}_2 \, {\rm cm}^{-2}$ to 
$1.65 \times 10^{26} \, {\rm H}_2 \, {\rm cm}^{-2}$).}
\label{fig:new2}
\end{figure*}

\begin{figure*}

\setlength{\unitlength}{1mm}
\begin{picture}(80,80)
\includegraphics{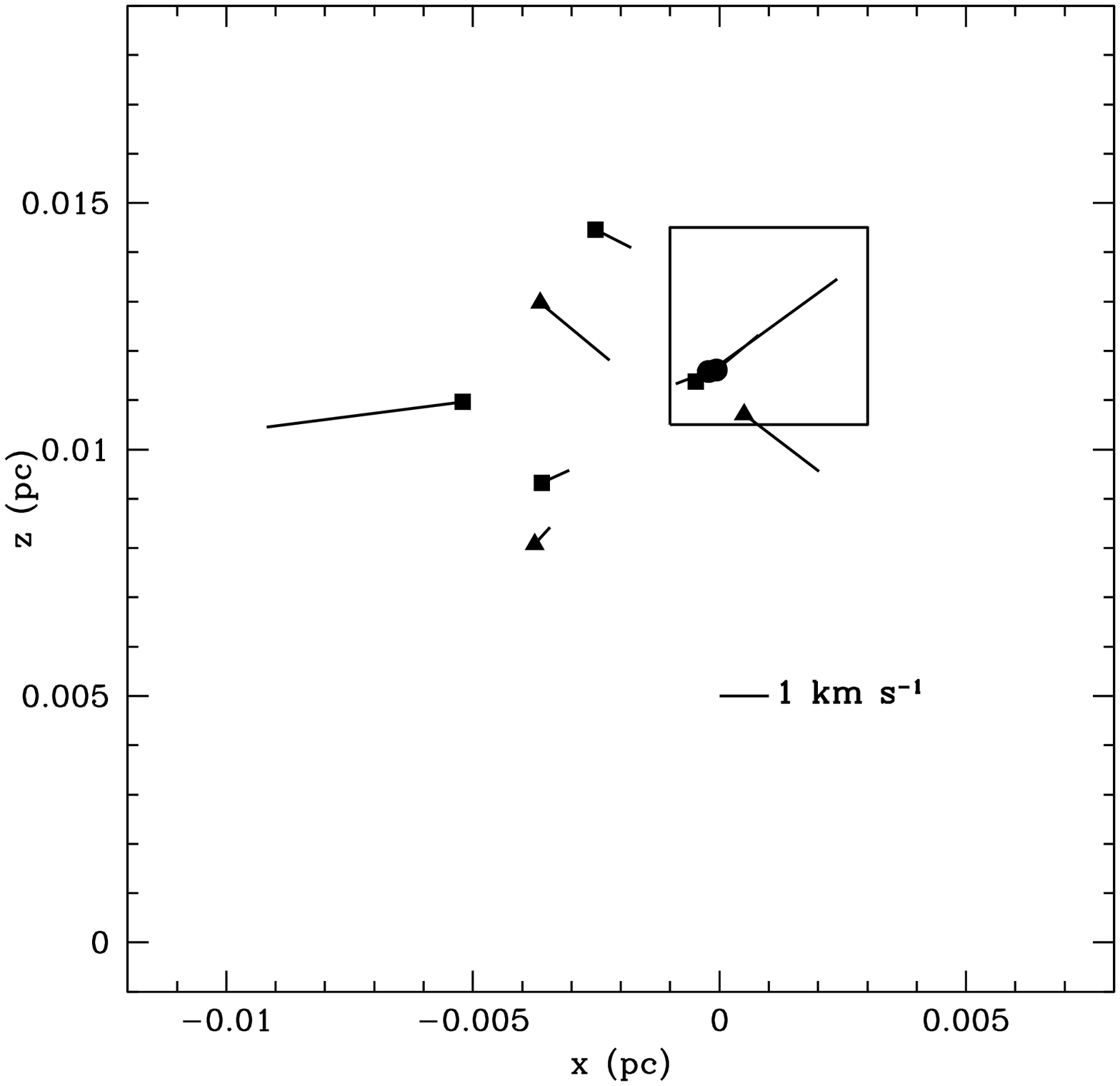}
\end{picture}
\begin{picture}(80,80)
\includegraphics{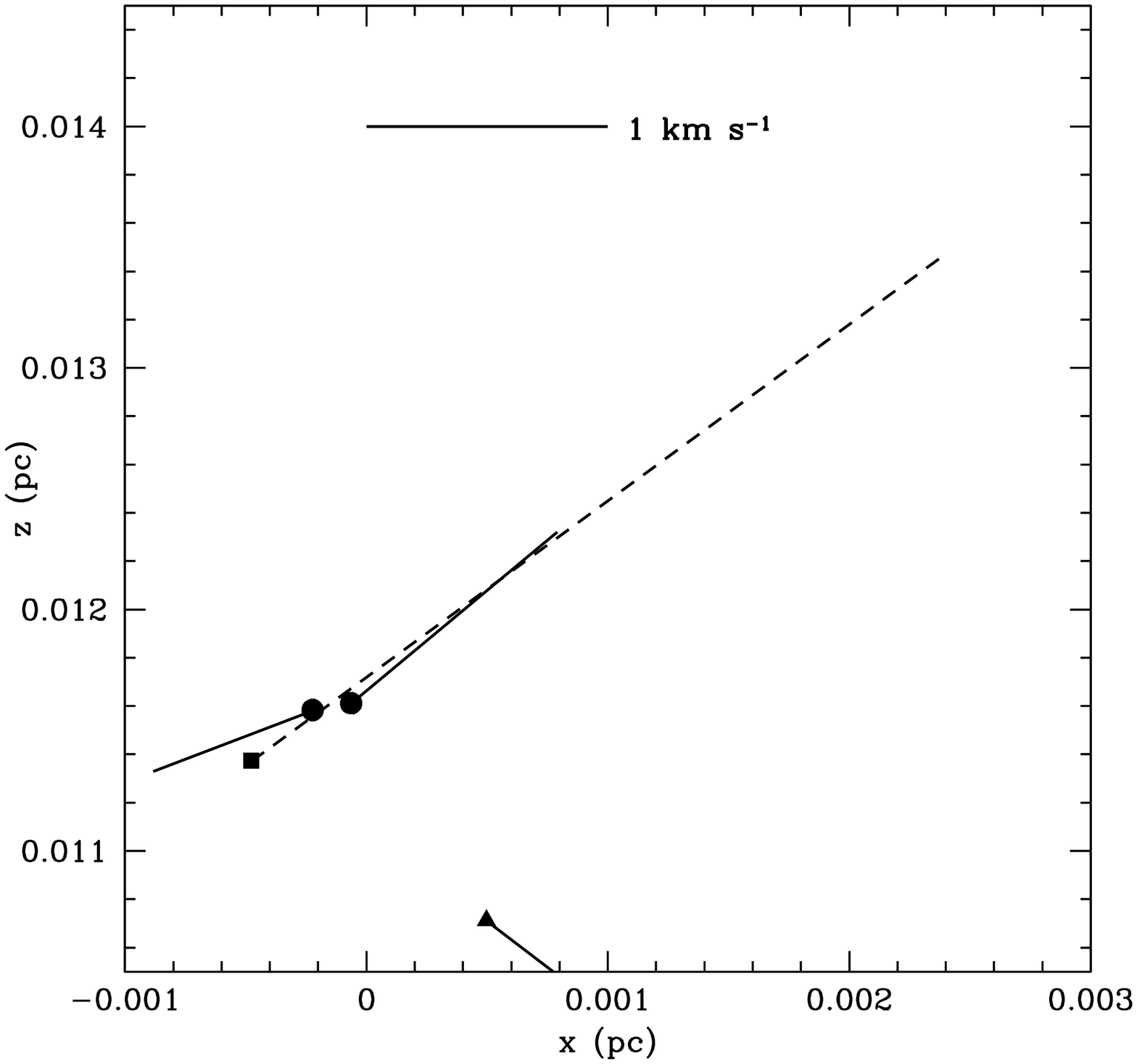}
\end{picture}

\caption{{\it Left Panel.} The positions and velocities of the 9 sinks shown 
in the right panel of Fig. \ref{fig:new2}, i.e. at the end of the simulation 
with sinks. Circles denote the binary components ($M>1 M_{\odot}$), squares 
denote low-mass protostars ($0.1 M_{\odot}<M<1 M_{\odot}$) and triangles 
denote protostars in the sub-stellar mass range ($M<0.1 M_{\odot}$). The lines 
indicate the projection of the velocity of each sink on the $xz$-plane 
using a scaling of 1000 km s$^{-1}$ : 1 pc. A 1 km s$^{-1}$ velocity line is 
indicated for clarity at the bottom of this panel. The direction of each 
velocity vector is pointing {\em away} from the corresponding symbols. 
{\it Right Panel.} Zooming in close to the binary (the region enclosed by the 
square in the left panel). For clarity the velocity of the low-mass companion 
of the binary is illustrated by a dashed line. The 1 km s$^{-1}$ velocity 
line is indicated at the top of this panel.}
\label{fig:new_vect}
\end{figure*}

\subsection{Repeating the $M_0 = 10 M_\odot$, $b = 0.2$, ${\cal M} = 10$ 
simulation with sink particles}{\label{sec:sinks}}

\subsubsection{The algorithm and sink parameters}

In order to derive a more direct estimate of the number and the final 
mass of the protostars that form in such collisions, we have repeated the 
$b = 0.2$, ${\cal M} = 10$ collision this time including sink particles 
\cite{BateMNRAS1995}. In particular, if certain conditions are satisfied 
(protostellar peak density exceeding a certain density threshold while the 
protostar is gravitationally bound; for details see below), we force all gas 
particles within a protostar to be removed from the simulation and replaced 
with a single collisionless (star) sink particle having as mass the sum of the 
masses of the gas particles it replaces\footnote{It is essential to introduce 
sinks automatically in the simulations, as the replacement by hand of 
protostars with sinks will lead the system to a new state of suspended 
animation as soon as another protostar forms, 
in which case we will have to replace this new protostar with another sink and 
so on.}. The position and the velocity of a sink are given 
by mass-weighted sums of the positions and the velocities, respectively, of 
the gas particles that the sink replaces. Subsequent to their formation, sinks 
are allowed to accrete more gas that enters their radius of influence, 
$r_{\rm sink}$. This is the only way the sinks interact with the gas, apart 
from the boundary conditions that they impose on their neighbouring gas 
particles (see the discussion below). 
Gas particles that are accreted by a sink are removed 
from the simulation. The sinks are subject only to the force of gravity, 
exchanging gravitational interactions with all types of particles. Sink-sink 
interactions are Plummer softened at distances smaller than 
$2 \times r_{\rm sink}$.

The conditions for sink creation are that a protostar is i) dense enough, i.e. 
its peak density exceeds a certain threshold, specifically $\rho_{\rm sink} = 
10^{-12}$ g cm$^{-3}$, which is 100 times higher than $\rho_1$, the 
density after which adiabatic heating switches on; and ii) gravitationally 
bound, i.e. the gas within the protostar 
has already started collapsing. In terms of the Jeans analysis, from the 
second of the above sink formation conditions we obtain the minimum mass of a 
sink that can be resolved at $\rho_{\rm sink}$ as well as the corresponding 
sink creation radius which also serves as the sink accretion radius, 
$r_{\rm sink}$. In particular, using the fact that for densities between 
$\rho_1 = 10^{-14}$ and $\rho_{\rm sink} = 10^{-12}$ g 
cm$^{-3}$ the gas evolves as $c \propto \rho^{1/3}$ ({\it cf}. Eqn. 
\ref{equa:statecool}) the minimum fragment mass at $\rho_1$ of Eqn. 
\ref{equa:jeans} translates to a minimum mass of a sink at $\rho_{\rm sink}$ 
of order $0.05 M_{\odot}$ for ${\cal N}_{\rm neib} \sim 100$ 
\cite{BateMNRAS1997}. Therefore, we obtain $r_{\rm sink} \sim 20$ AU. 

Gas particles are accreted by a sink when i) they have approached the sink at 
a distance less than $r_{\rm sink}$, and ii) they are gravitationally bound to 
the sink. In order to model accretion by sinks 
properly, a number of boundary corrections must be applied to the neighbouring 
gas particles of a sink, so that these gas particles do not feel a 
discontinuity in the density field 
due to the proximity of a sink.
We have implemented all the boundary corrections given by Bate 
{\it et al}. \shortcite{BateMNRAS1995}.

\subsubsection{Results}

The two protostars identified at the end of the simulation {\it without} 
sinks (Fig. \ref{fig:b2m10xy}), are replaced by sinks at $t \sim 0.391$ and 
$t \sim 0.393$ Myr (upper left panel of Fig. \ref{fig:new1}; these are the 
sinks on the right and in the middle of the panel, respectively). Note that 
the sink on the right hand side of the upper left panel of Fig. \ref{fig:new1}
lies below (lower $y$ value) the sink in the middle of the panel ({\it cf}. 
the left panel of Fig. \ref{fig:b2m10xy}). As the simulation with sinks 
progresses, the sink on the right hand side of the upper left panel of Fig. 
\ref{fig:new1} first moves to the left towards the centre of the computational 
domain and later it is forced to move along the $z$-axis by shear produced by 
material entering the shock compressed layer (at an angle) from the 
opposite side. At the time of the upper right panel of Fig. \ref{fig:new1}, 
this sink has exited the small field of view of this panel (from the top side).

In the meantime, a third protostar formed ({\it cf}. Section \ref{sec:b2m10}) 
in the top left hand corner of the displayed region and at $t \sim 0.400$ Myr 
it was replaced by a sink. At the time of the upper right panel of Fig. 
\ref{fig:new1} ($t \sim 0.402$ Myr), the two sinks remaining in the displayed 
region have already approached each other along the filament. They have a 
close encounter with periastron $\sim 200$ AU at $t \sim 0.403$ Myr and 
capture themselves into a binary. A large circumbinary disc forms around the 
binary. The binary components continue accreting matter from this disc and 
their orbit is hardened. At the same time, the disc material is 
replenished by material falling on it from the filaments. Interactions between 
the accretion flows and the disc lead to the formation of further companions 
to the binary (see e.g. the sinks on the spiral arms of the disc in the lower 
panels of Fig. \ref{fig:new1}). Most of these companions are ejected within a 
few orbits around the binary after having a close encounter with it (e.g the 
sinks in the lower density regions of the left panel of Fig. \ref{fig:new2}).

At the end of the simulation with sinks (right panel of Fig. \ref{fig:new2}, 
$t_{\rm evol} \sim 2.6 \; t_{\rm ff}$), the binary components have mass of 
$\sim 1.5 M_{\odot}$ each, and their separation is $\sim$ 40 AU, i.e. the 
minimum separation resolved due to the Plummer softening that is used to model 
sink-sink interactions. Eight additional sinks have formed in total (including 
a sink that has exited the displayed region; this sink has final mass of 
$\sim 0.5 M_{\odot}$). The masses of the secondary protostars, grouped in two 
mass bins, are denoted by different symbols in Fig. \ref{fig:new_vect} 
(squares for $0.1 M_{\odot}<M<1.0 M_{\odot}$ and triangles for 
$M<0.1 M_{\odot}$). Their projected velocities are also indicated by the lines 
in Fig. \ref{fig:new_vect}. None of them remains bound to the binary. However, 
they have been ejected from the system with moderate velocities of order a few 
km s$^{-1}$ (e.g. the object having a close encounter with the binary on the 
right panel of Fig. \ref{fig:new_vect}).

A more detailed account of the properties of this system of protostars is 
beyond the scope of this paper. We note, however, that the 
general picture drawn from our simulation with sinks with respect to the 
formation and evolution of this system of protostars is similar to that of one 
of the subclusters reported by Bate \& Bonnell \shortcite{BateMNRAS2005}, 
which has also formed along filaments. A small difference to the results of 
Bate \& Bonnell is the lower rate with which secondaries form in our 
simulation, which we attribute to the stiffer equation of state used here 
(with polytropic exponent of $5/3$ instead of $7/5$ that is used by Bate \& 
Bonnell). A stiff equation of state 
provides proto-fragments with larger thermal pressure support. For example, we 
have noted that fragments like the one in the top left hand corner of the 
right panel of Fig. \ref{fig:new2} form a sink only when they fall onto the 
disc, i.e. when they get very quickly loaded with mass exceeding a Jeans mass. 
We also note that no individual discs and/or envelopes can be resolved 
here for any of the 10 sinks formed, i.e. if any of these protostars is 
attended by a disc and/or surrounded by an envelope, this disc/envelope will 
be smaller than $r_{\rm sink} \sim 20$ AU in size. The separation of the 
binary is, accordingly, expected to be smaller than $\sim 40$ AU which is the 
minimum separation than can be resolved. Finally, we note that, despite the 
very small number statistics, the mass distribution of the sinks formed here 
is consistent with the observed protostellar mass distributions in several 
star forming regions.

The total mass in sinks at the end of the simulation is 5.1 $M_{\odot}$, which 
implies a SFE of order 25\%. There is, however, about 3 $M_{\odot}$ more bound 
gas in the filaments, but we expect only the binary components to have access 
to this material. Moreover, we expect accretion onto these two protostars to 
start decreasing at subsequent times, i.e. when they enter the Class I 
phase, during which stellar feedback effects (which we do not model here) 
start affecting the disc from which they now accrete. We do not anticipate, 
therefore, more than 1-2 $M_{\odot}$ of additional stellar mass in the binary 
and/or additional secondaries, limiting the SFE of this collision to a maximum 
value of $\sim$ 30\%. Because this collision is one of the most efficient 
among the collisions we study here, we estimate the average SFE of the 
low-mass clump collisions to be of order 20-30\%. After having conducted the 
simulation with sinks, we note that, in terms of 
primary protostellar mass, our initial SFE estimate, which was made in the 
previous section by extrapolating the masses obtained from the simulations 
without sinks and using well known protostellar mass accretion rates and 
lifetimes, was rather accurate. The SFE estimate obtained from the simulation 
with sinks also accounts for the formation of secondary protostars, which we 
could not follow before introducing sinks, and has lead to this higher SFE of 
order 20-30\%.

\section{$75 M_\odot$ clump collisions}{\label{sec:amar}}

In this section we present the results of the suite of three simulations 
involving collisions between clumps having mass $M_0 = 75 M_\odot$. The 
Mach Number is set to ${\cal M} = 9$, and the offset parameter is set to 
$b = 0.2$, $0.4$, and $0.5$. The purpose of this suite of simulations 
is to explore how the results depend on clump mass $M_0$, and to test the 
reliability of the results reported by Bhattal {\it et al}. 
\shortcite{BhattalMNRAS1998}.

The main effect of increasing the clump mass, $M_0$, and hence also the 
clump radius, $R_0$, is that -- at fixed $b$ -- the shock compressed layer is 
more extended and the fragmentation scale is also somewhat greater 
({\it cf}. Eqn. \ref{equa:fragmentsep}). Consequently, the 
shock compressed layer breaks up into a more complex network of filaments, as 
the number of filaments should be $\propto M_{0}^{1/4}$.

The main conclusion concerning the simulations reported by Bhattal 
{\it et al}. \shortcite{BhattalMNRAS1998} is that their results are seriously 
corrupted, both by the fact that the Jeans condition is violated, and by the 
fixed (and rather large) gravity softening length.

\begin{figure*}

\setlength{\unitlength}{1mm}
\begin{picture}(80,90)
\includegraphics{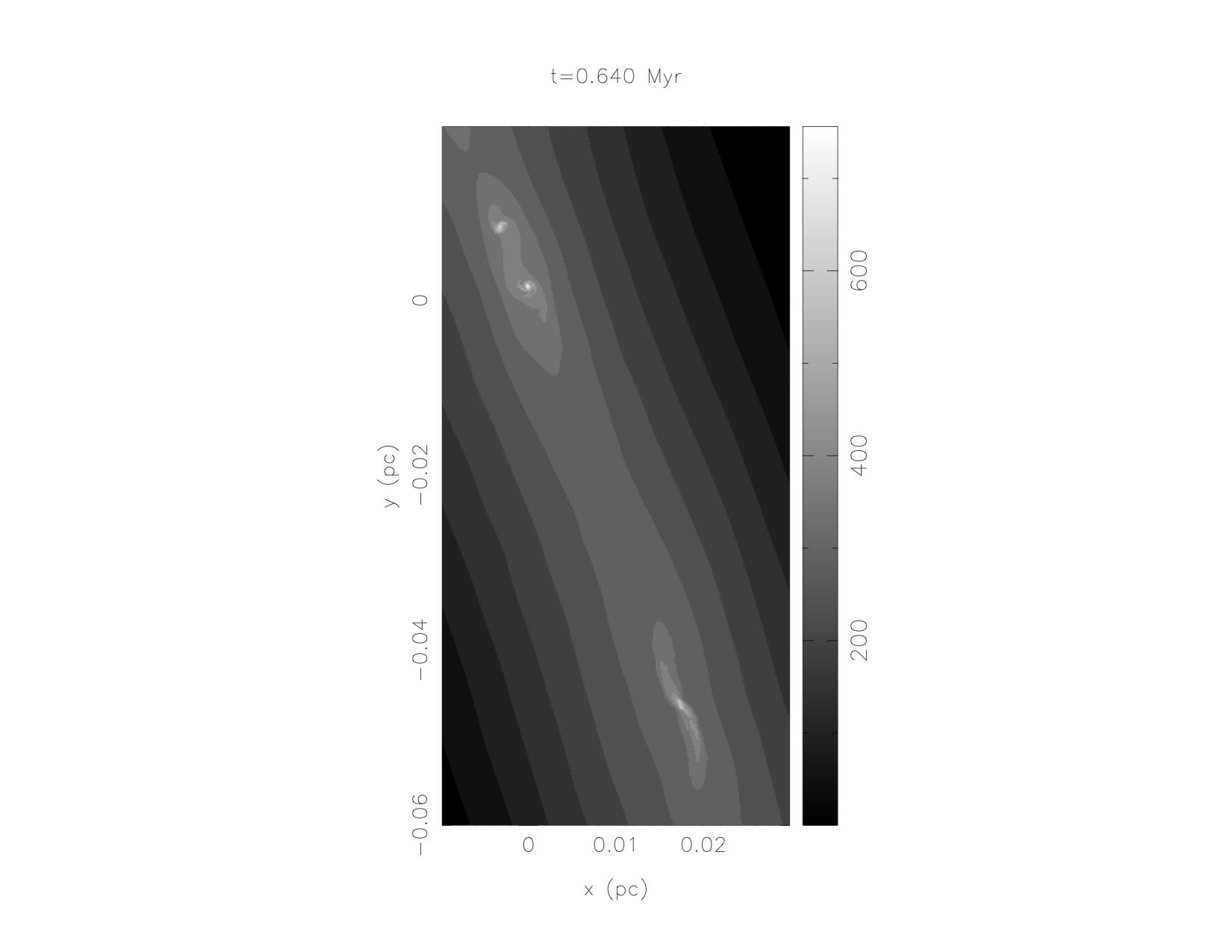}
\end{picture}
\begin{picture}(80,90)
\includegraphics{kitsionas_fig12b.ps}
\end{picture}

\caption{$M_0 = 75 M_\odot$, $b = 0.2$, ${\cal M} = 9$, at the end 
of the simulation, ($t = 0.640\,{\rm Myr}$). 
{\it Left Panel.} View along the $z$-axis; $\Delta x = 0.04\,{\rm pc}$, 
$\Delta y = 0.08\,{\rm pc}$; sixteen-interval logarithmic grey-scale, 
in units of g cm$^{-2}$, from 
$3.63 \times 10^{-2} \, {\rm g} \, {\rm cm}^{-2}$ to 
$7.59 \times 10^{2} \, {\rm g} \, {\rm cm}^{-2}$ 
($9.08 \times 10^{21} \, {\rm H}_2 \, {\rm cm}^{-2}$ to 
$1.90 \times 10^{26} \, {\rm H}_2 \, {\rm cm}^{-2}$). 
{\it Right Panel.} View along the $y$-axis; $\Delta x = \Delta z = 
0.028\,{\rm pc}$; sixteen-interval logarithmic grey-scale, 
in units of g cm$^{-2}$, from 
$2.00 \times 10^{-1} \, {\rm g} \, {\rm cm}^{-2}$ to 
$1.70 \times 10^{3} \, {\rm g} \, {\rm cm}^{-2}$ 
($5.00 \times 10^{22} \, {\rm H}_2 \, {\rm cm}^{-2}$ to 
$4.25 \times 10^{26} \, {\rm H}_2 \, {\rm cm}^{-2}$).}
\label{fig:amar2}
\end{figure*}

\begin{figure}

\setlength{\unitlength}{1mm}
\begin{picture}(80,90)
\includegraphics{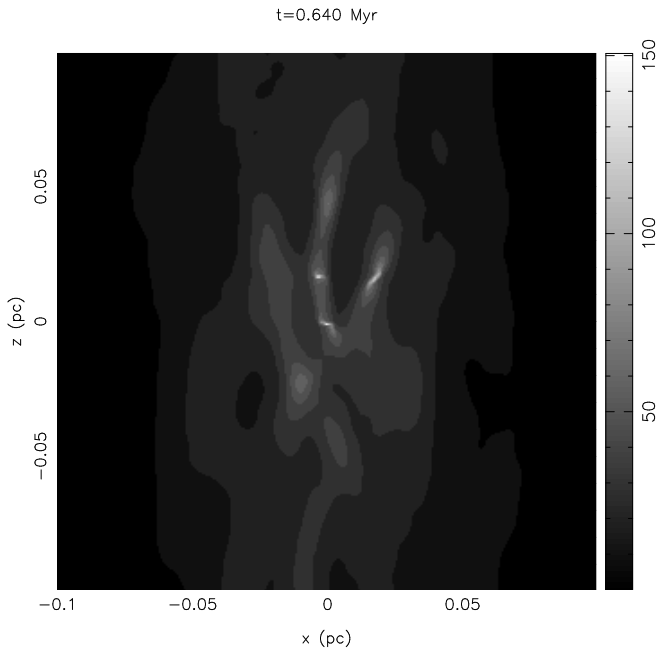}
\end{picture}

\caption{$M_0 = 75 M_\odot$, $b = 0.2$, ${\cal M} = 9$, at the end 
of the simulation, ($t = 0.640\,{\rm Myr}$), viewed along the $y$-axis 
to show the network of filaments produced; $\Delta x = \Delta z = 
0.2\,{\rm pc}$; sixteen-interval logarithmic grey-scale, 
in units of g cm$^{-2}$, from 
$7.24 \times 10^{-2} \, {\rm g} \, {\rm cm}^{-2}$ to 
$1.51 \times 10^{2} \, {\rm g} \, {\rm cm}^{-2}$ 
($1.81 \times 10^{22} \, {\rm H}_2 \, {\rm cm}^{-2}$ to 
$3.78 \times 10^{25} \, {\rm H}_2 \, {\rm cm}^{-2}$).}
\label{fig:amar2fil}
\end{figure}

\subsection{$M_0 = 75 M_{\odot}$, $b = 0.2$, ${\cal M} = 9$}{\label{sec:amar2}}

After $\sim 0.61\,{\rm Myr}$, the shock compressed layer has become 
sufficiently dense and massive to break up into a network of tumbling 
filaments. Protostars then start to condense out at the intercepts of the 
filaments. On-the-Fly splitting starts at $\sim 0.63\,{\rm Myr}$. In total, 
four protostars form, but two of them quickly merge. The three surviving 
protostars are rotating rapidly, and attended by accretion discs with 
strong spiral structure. Because of the increasing offset between the opposing 
streams accreting onto a protostar along the filament, the protostellar disc 
is spun up, and we anticipate that the disc will eventually fragment to 
produce secondary companions. However, the simulation has been terminated 
before this happens. The accretion streams are very smooth.

In comparison with the lower-mass ($10 M_\odot$) collision at $b = 0.2$, 
more filaments and more protostars are formed, because the area of the 
shock compressed layer is larger and the fragmentation scale $L_{\rm frag}$ 
is only moderately larger.

At the end of the simulation ($\sim 0.64\,{\rm Myr}$; Figs. \ref{fig:amar2} 
\& \ref{fig:amar2fil}), the total mass of the 
three protostars is $\sim 0.85 M_\odot$, and their radii are in the range 
$100\;\,{\rm to}\;130\,{\rm AU}$. Their central densities exceed $\rho_1$, 
so they have already started to heat up adiabatically. The separation between 
the protostars is $\sim 10^4\,{\rm AU}$ (Fig. \ref{fig:amar2}), and they are 
only weakly bound, so it is almost certain that within 1 Myr one of them will 
be ejected gravitationally by the other two. The minimum density in the 
filaments is $\rho_{\rm fil} \sim 3.0 \times 10^{-17}\,{\rm g}\,{\rm cm}^{-3}$ 
($n_{{\rm H}_{2}} \sim 5 \times 10^{6}\,{\rm cm}^{-3}$, Fig. 
\ref{fig:amar2fil}). The number of active particles has increased from 220,000 
to 260,000.

In contrast, Bhattal {\it et al}. \shortcite{BhattalMNRAS1998} found that for 
low-$b$ collisions a single primary protostar formed at the centre of the 
collision and accreted material from a single filament. The primary protostar 
rapidly acquired an accretion disc, and grew in mass and angular momentum as 
the offset between the opposing accretion streams along the filament 
increased. Due to the development of spiral structure in the disc, and the 
lumpiness of the accretion streams, lower-mass secondary companions formed in 
the disc around the primary. There was no suggestion of the shock compressed 
layer fragmenting into multiple filaments. Moreover, the secondary companions 
modelled by Bhattal {\it et al}. were not always 
resolved properly, as in some cases they contained fewer than 50 particles. 
Also Bhattal {\it et al}. used a large and constant gravity softening 
length $\epsilon \sim$500 AU, so that they could not resolve the formation 
of an object until its size became $\gtrsim$500 AU. 

We conclude that the evolution of the density field is followed more 
faithfully in our current simulations with Particle Splitting. The fact 
that the Jeans condition is obeyed, prevents a central object from 
forming artificially before the filaments. Also there is no preferred 
length scale, so we can trust the detailed evolution of the discs and 
their dynamical interaction with the filaments.

\begin{figure}

\setlength{\unitlength}{1mm}
\begin{picture}(80,90)
\includegraphics{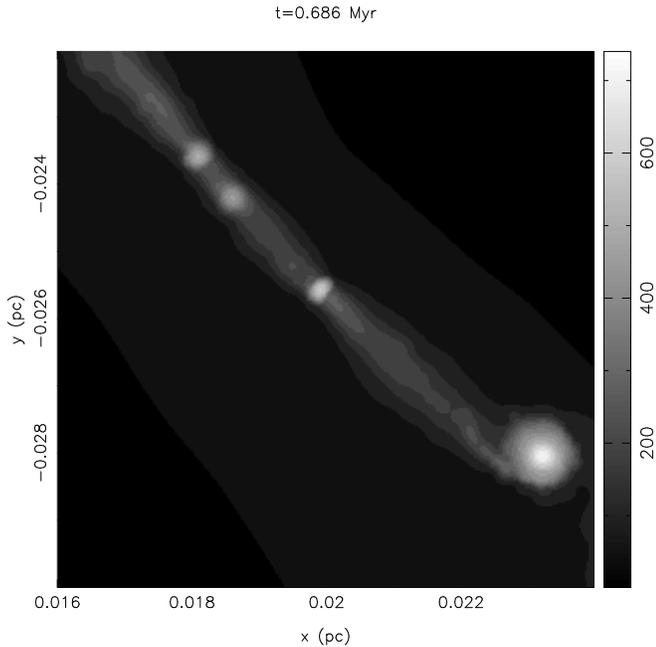}
\end{picture}

\caption{$M_0 = 75 M_\odot$, $b = 0.4$, ${\cal M} = 9$, at the end of the 
simulation ($t = 0.686\,{\rm Myr}$), viewed along the $z$-axis; $\Delta x = 
\Delta y = 0.008\,{\rm pc}$; sixteen-interval logarithmic grey-scale, 
in units of g cm$^{-2}$, from 
$9.33 \times 10^{-1} \, {\rm g} \, {\rm cm}^{-2}$ to 
$7.41 \times 10^{2} \, {\rm g} \, {\rm cm}^{-2}$ 
($2.33 \times 10^{23} \, {\rm H}_2 \, {\rm cm}^{-2}$ to 
$1.85 \times 10^{26} \, {\rm H}_2 \, {\rm cm}^{-2}$).}
\label{fig:amar4}
\end{figure}

\subsection{$M_{0} = 75 M_\odot$, $b = 0.4$, ${\cal M} = 9$}{\label{sec:amar4}}

A single tumbling filament forms at $\sim 0.66\,{\rm Myr}$. On-the-Fly 
Particle Splitting starts at about the same time. At least four protostars 
condense out of the filament; a further two may have started to condense out. 

At the end of the simulation ($\sim 0.686\,{\rm Myr}$; Fig. \ref{fig:amar4}), 
the total mass of the four well established protostars is $\sim 0.40 
M_{\odot}$. They are positioned randomly along the filament, and they are 
falling towards one another. The most massive protostar is disc-like with a 
radius of $\sim 170\,{\rm AU}$ (Fig. \ref{fig:amar4}). The three smaller 
protostars are still roughly spherical with radii $\sim 45\,{\rm AU}$. 
Their central densities exceed $\rho_1$, so they have started heating up 
adiabatically. The number of active particles has increased from 220,000 to 
260,000.

This is very similar to what happened in the corresponding simulation 
performed by Bhattal {\it et al}. \shortcite{BhattalMNRAS1998}, except that 
their simulation was followed further, and the disc round the most massive 
protostar was therefore spun up by 
accretion along the tumbling filament. As a consequence it became 
rotationally unstable, and fragmented to produce secondary companions. 
This might also have happened here, if we had been able to follow the 
simulation for long enough. However, once again we can have little faith in 
the Bhattal {\it et al}. result because some of the secondaries contained at 
their inception less than 50 particles, and gravity was severely 
softened on scales $\lesssim 500\,{\rm AU}$.

\begin{figure}

\setlength{\unitlength}{1mm}
\begin{picture}(80,90)
\includegraphics{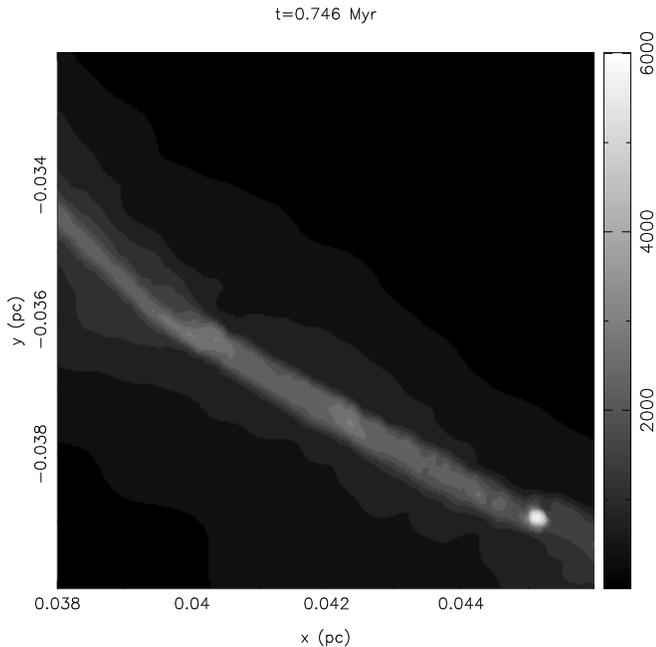}
\end{picture}

\caption{$M_0 = 75 M_\odot$, $b = 0.5$, ${\cal M} = 9$ at the end of the 
simulation ($t = 0.750\,{\rm Myr}$), viewed along the $z$-axis; $\Delta 
x = \Delta y = 0.008\,{\rm pc}$; sixteen-intervals logarithmic grey-scale, 
in units of g cm$^{-2}$, from 
$3.98 \times 10^{-1} \, {\rm g} \, {\rm cm}^{-2}$ to 
$6.03 \times 10^{3} \, {\rm g} \, {\rm cm}^{-2}$ 
($9.95 \times 10^{22} \, {\rm H}_2 \, {\rm cm}^{-2}$ to 
$1.51 \times 10^{27} \, {\rm H}_2 \, {\rm cm}^{-2}$).}
\label{fig:amar5}
\end{figure}

\subsection{$M_{0} = 75 M_\odot$, $b = 0.5$, ${\cal M} = 9$}{\label{sec:amar5}}

On-the-Fly Particle Splitting starts at $\sim 0.72\,{\rm Myr}$. At 
$\sim 0.73\,{\rm Myr}$, a single tumbling filament forms, and fragments 
into several protostars. Two of them, at the bottom right hand corner of 
Fig. \ref{fig:amar5}, condense out first and eventually, at 
$\sim 0.746\,{\rm Myr}$, they merge. Subsequently, another three protostars 
form, two near the centre of the filament and one at 
the left hand end (Fig. \ref{fig:amar5}). All fragments remain spherical and 
are strongly centrally condensed. Some other density enhancements appear in 
the filament, suggesting that more fragments may be forming.

At the end of the simulation ($\sim 0.75\,{\rm Myr}$), the total mass of the 
four well established protostars is $\sim 0.65 M_{\odot}$. The most massive 
one has radius $\sim 35\,{\rm AU}$ and the radii of the other three are 
$\sim 100\,{\rm AU}$. Their central densities exceed $\rho_1$, so they have 
started to heat up adiabatically. The minimum density in the filament is 
$\rho_{\rm fil} \sim 2 \times 10^{-18}\,{\rm g}\,{\rm cm}^{-3}$ 
($n_{{\rm H}_{2}} \sim 5 \times 10^{5}\,{\rm cm}^{-3}$). The number of 
active particles has increased from 220,000 to 342,000.

Again, this is very similar to what happened in the corresponding simulation 
performed by Bhattal {\it et al}. \shortcite{BhattalMNRAS1998}, where a single 
tumbling filament formed and then fragmented. Bhattal {\it et al}. performed a 
standard resolution simulation of this case, which produced two protostars 
after $\sim 1\,{\rm Myr}$, and a higher-resolution simulation in which each 
component of the binary was itself a binary, i.e. an hierarchical quadruple. 

\subsection{Discussion of $75 M_\odot$ collisions}{\label{sec:amar-disc}}

In the $b = 0.2$ collision, the shock compressed layer has considerable 
lateral extent, and fragments into a network of tumbling filaments; multiple 
protostars then condense out of these filaments. In the $b = 0.4$ and 
$b = 0.5$ collisions, the extent of the shock compressed layer is 
considerably smaller, and in the first instance it produces a single 
tumbling filament, which then fragments into a line of protostars. As 
these protostars accrete material with increasing specific angular momentum 
from the tumbling filament, accretion discs form around them and develop 
spiral structure. At the same time the protostars fall towards one another, 
and they are therefore likely to interact, either merging, being captured into 
binary systems, or triggering the formation of further protostars.

The main difference between the $10 M_\odot$ collisions of Section 
\ref{sec:low-mass}, and the $75 M_\odot$ collisions of this section, is that 
the latter involve more mass {\it and} produce a shock compressed layer of 
greater extent. Consequently the filaments are longer and spawn more 
protostars, typically 4 to 6 (as compared with 1 to 2 for the $10 M_\odot$ 
collisions). Despite the fragmentation scale being somewhat larger, there 
tend to be more filaments, at least for low-$b$ collisions.

Taking into account that in the $75 M_\odot$ collisions i) there are on 
average three times more primary protostars than in the $10 M_\odot$ 
collisions, and ii) there is more mass available for protostars to accrete in 
the shock compressed layer (and thus also in the filaments), and using the SFE 
estimated in Section \ref{sec:sinks} for the $10 M_\odot$ collisions, we 
arrive to an average SFE estimate of order 10-15\% for the $75 M_\odot$ 
collisions (as the total available gas mass is 7.5 times larger).

It seems inescapable that the results obtained by Bhattal {\it et al}. 
\shortcite{BhattalMNRAS1998} were corrupted, both by the very large gravity 
smoothing length, and by the fact that the Jeans condition was violated (i.e. 
the Jeans mass was not resolved at all times). The inferences which they made 
are therefore not reliable.

\section{Conclusions}{\label{sec:conclusi}}

We have conducted a series of simulations of star formation triggered by 
low-velocity collisions between low-mass molecular clumps. By implementing 
On-the-Fly Particle Splitting in these simulations, we have ensured that the 
Jeans condition is satisfied throughout the computational domain, at all 
times. Consequently we are confident that our simulations have not been 
corrupted by artificial fragmentation, and that real fragmentation has not 
been inhibited by inadequate resolution. We are therefore now in a position to 
apply our SPH code with Particle Splitting to simulations of clump-clump 
collisions with more realistic initial conditions, for example clumps of 
unequal mass, clumps with internal turbulence, and/or rotating clumps. 
The use of Particle Splitting will afford significant advantages in terms 
of computational efficiency. 

In order 
to confirm the usefulness and reliability of this technique, we have 
repeated the $M_0 = 10 M_\odot$, $b = 0.2$, ${\cal M} = 10$ simulation, 
using a standard code without Particle Splitting, but introducing 
less massive (13 times less massive) particles, from the outset, and 
13 times as many of them (i.e. 390,000 particles in total). The results 
differ only in small details, not in the overall statistics of fragmentation, 
such as the number and mass of the protostars formed. 
Moreover, the standard simulation with 390,000 particles uses $\sim\,6$ 
times more memory, and $\sim\,2$ times more CPU than the simulation using 
On-the-Fly Particle Splitting with only 30,000 particles initially.

The simulations presented here demonstrate that the 
primary mode creating protostars, following a clump-clump collision, entails 
the formation of a shock compressed layer (while the gas is cooling 
radiatively), its gravitational fragmentation 
into one or more filaments (the number depending on the lateral extent of 
the layer and the fragmentation scale), the break up of the filaments into 
cores, the condensation of primary protostars out of the cores, and the 
subsequent growth of the protostars by accretion along the filaments. Similar 
filamentary structures and cores have been found in the simulations of 
Pongracic {\it et al}. \shortcite{PongracicMNRAS1992}, Klessen \& Burkert 
\shortcite{KlessenApJSS2000,KlessenApJ2001a}, Klessen 
\shortcite{KlessenApJ2001b}, Bate {\it et al}. 
\shortcite{BateMNRAS2002a,BateMNRAS2002b,BateMNRAS2003}, Bate \& Bonnell 
\shortcite{BateMNRAS2005}, Jappsen {\it et al}. \shortcite{JappsenAnA2005}, 
Martel, Evans \& Shapiro \shortcite{MartelApJSS2006}.

Increasing the clump mass $M_0$ (with $b$ and ${\cal M}$ held constant) 
increases the number of primary protostars, because the mass and lateral 
extent of the shock compressed layer is greater, and the fragmentation scale 
$L_{\rm frag}$ is only moderately larger. Increasing $b$ (with $M_0$ and 
${\cal M}$ held 
constant) reduces the number of primary protostars, because the lateral extent 
of the shock compressed layer is reduced (and the fragmentation length is 
increased somewhat by shear). Increasing $b$ also increases the orbital 
angular momentum of the resulting ensemble of protostars. Increasing 
${\cal M}$ (with $M_0$ and $b$ held constant) increases the number of primary 
protostars because it reduces $L_{\rm frag}$. It also increases the orbital 
angular momentum of the resulting ensemble of protostars.

For collisions with $b \gtrsim 0.6$, no significant shock compressed layer 
forms, and the clump-clump collision does not trigger star formation. For 
collisions with $b \lesssim 0.5$, the filaments and cores that form have 
densities $n_{{\rm H}_2} \gtrsim 10^5\,{\rm cm}^{-3}$. Therefore it should be 
possible to map them in NH$_3$ or CS, in nearby star formation regions. The 
protostars which condense out of a core have much higher central densities 
$n_{{\rm H}_2} \gtrsim 10^{10}\,{\rm cm}^{-3}$. Material accretes onto the 
protostars at rates in the range $1\;{\rm to}\;5 \times 10^{-5} 
M_\odot\,{\rm yr}^{-1}$, and this lasts for a time interval in the range 
$1\;{\rm to}\;3 \times 10^4\,{\rm yrs}$. Therefore we should identify these 
protostars with the Class 0 phase of evolution \cite{LadaLADA1999}.

Hartmann \shortcite{HartmannApJ2002} has recently pointed out that young stars 
in Taurus form primarily within gaseous filaments, via fragmentation of the 
filaments; that the mean separation between the young stars in Taurus is of 
order the local Jeans length; and that the protostellar cores are elongated 
along the filaments. This is very similar to the phenomenology which we 
observe in our simulations. In our simulations, the shock compressed layer 
created by the clump-clump collision fragments into filaments, and the 
filaments then fragment into cores; the separations between cores are 
of order the local Jeans length; and the individual cores are prolate 
and elongated parallel to the filaments. Our findings on filament 
fragmentation derived from clump-clump collision simulations agree with those 
of Jappsen {\it et al}. \shortcite{JappsenAnA2005} obtained from simulations 
of the turbulent ISM. They also fit the theoretical predictions of Larson 
\shortcite{LarsonMNRAS2005} on the fragmentation of filaments and the 
influence of thermodynamics.

The prolateness of the cores in our 
simulations is only evident from the grey-scale plots we have presented, 
if one takes into account that molecular-line observations tend 
to be most sensitive to gas at the corresponding critical density 
($\sim 10^3\,{\rm cm}^{-3}$ to $\sim 10^5\,{\rm cm}^{-3}$, for the lines used 
to map cores in Taurus), and submillimeter observations are optically thick 
for column-densities exceeding $4 \times 10^{25}\,{\rm H}_2\,{\rm cm}^{-2}$. 
In other words one must look at the grey contours to see the structures which 
would be revealed by molecular-line or submillimeter continuum observations. 
Therefore, whereas the protostellar discs formed in our simulations present 
a range of shapes (dependent on whether they are viewed close to edge-on or 
close to pole-on), the cores within which these discs are embedded are almost 
invariably prolate and elongated along the filament from which they are 
condensing. The elongation along the filament arises because the cores are 
usually being fed by two opposing accretion streams from along the filament.

Because the filaments are tumbling, the accretion flow onto a protostar 
tends to deliver ever increasing specific angular momentum, and therefore 
the protostar is spun up. This may lead to rotational instabilities which 
create secondary companions to the primary protostars. However, the 
simulations have been terminated (due to limited computational resource) 
before there is any clear evidence of this happening.

Additionally, protostars condensing out of the same filament will often 
fall together, and therefore they may interact. Ultimately, this could 
result in a merger, or in capture to form a binary, or in further 
fragmentation to produce secondary protostars, but again we have not 
followed the simulations for that long. The temptation is to introduce sink 
particles, but then the interaction is not properly 
modelled, because of the over-simplifications associated with the 
introduction of and the interaction between sink particles. Nevertheless, we 
have repeated one of the low-mass clump collision simulations using sink 
particles. This attempt has shown that, indeed, pairs of primary protostars 
can form close binaries through capture, as they move towards each other along 
a tumbling filament. It has also been shown that such binaries can be attended 
by massive circumbinary discs and that the formation of secondary companions 
to such a binary is possible through the interaction of its disc with the 
accretion flows onto it from along the filament.

The SFE of low-mass clump collisions is estimated to be of order 20-30\% based 
on the result of the simulation with sinks. We have asserted that the SFE 
decreases with increasing mass and, in particular, that the SFE of the 
$75 M_\odot$ collisions is of order 10-15\%. Both these values are in 
accordance with the findings of Hunter {\it et al}. \shortcite{HunterApJ1986} 
from simulations of colliding gas flows. The SFEs estimated from our 
simulations are also consistent with those (20-30\%) observed in a 
number of molecular clouds \cite{RengarajanApJ1984}, which implies that 
clump-clump collisions can account, at least partly, for the star formation in 
those molecular clouds that can provide this triggering mechanism. 

Taken at face value, the above (theoretical and observational) 
estimates for the SFE could be interpreted in the following sense: in the 
absence of any other triggering mechanism, star formation in the ISM can 
be attributed mainly to low-mass clump collisions and not to collisions 
between higher-mass clumps. Evidently, this is not true in nature: first, 
these SFE estimates are only meaningful in a statistical sense, i.e. 
there is great variation from region to region; second, there is a number 
of complementary and/or mutually excluding mechanisms that are 
simultaneously in play. Collisions between low-mass clumps (and/or 
small-scale shocks produced in the turbulent ISM as part of larger-scale 
colliding flows) appear to be one such mechanism able to explain part of 
the star formation observed in the Galaxy.

Further progress will probably also necessitate simulations at even higher 
resolution, e.g. by employing multiple levels of particle splitting and using 
smaller sink particles (Kitsionas {\it et al}., in prep.). The detailed study 
of discs will also require the 
introduction of a more sophisticated algorithm to regulate the shear 
viscosity, for instance by using the time-dependent formulation of Morris \& 
Monaghan \shortcite{MorrisJCP1997} and/or the Balsara 
\shortcite{BalsaraJCP1995} switch, the implementation of a realistic 
continuity equation advocated by Imaeda \& Inutsuka \shortcite{ImaedaApJ2002}
(despite the recent criticism by Monaghan \shortcite{MonaghanMNRAS2006} of the 
need of the latter formalism), and possibly all these should be attempted in 
combination with a different method for the velocity calculation, such as XSPH 
\cite{MonaghanMNRAS2002}.

\section*{Acknowledgements}

The authors thank the anonymous first referee for useful and constructive 
comments and especially for his/her suggesting of the implementation 
of sink particles in our SPH code with particle splitting. They also 
thank the second referee, Ralf Klessen, for his comments and suggestions as 
well as Simon Glover for proof reading the manuscript.
The authors are thankful to Simon Goodwin, Henri Boffin and Neil Francis 
for many interesting discussions on this project as well as for their 
assistance on the implementation of sinks. They acknowledge the use of 
the Sun E4000 computer of the Cardiff Centre for Computational Science and 
Engineering as well as Sun workstations at both the School of Physics 
\& Astronomy, Cardiff University and the Institute of Astronomy \& 
Astrophysics, National Observatory of Athens. They also acknowledge the 
use of the public visualisation tool ``SuperSPHplot'' developed by Daniel 
Price. For part of 
the work presented in this paper, SK kindly acknowledges support 
by an EU Commission ``Marie Curie Intra-European (Individual) Fellowship'' 
of the 6th Framework Programme. Finally, SK would like to dedicate this 
paper to the memory of his school-time friend Dimitrios Velenis. 

\bibliography{kitsionas}

\label{lastpage}

\end{document}